\documentclass[sn-mathphys]{sn-jnl}

\jyear{2021}%
\usepackage{cancel}
\usepackage{comment}

\theoremstyle{thmstyleone}%
\theoremstyle{thmstyletwo}%

\theoremstyle{thmstylethree}%

\begin{document}

\title[Charting mobility patterns in the scientific knowledge landscape]{Charting mobility patterns in the scientific knowledge landscape}

\author[1,2]{\fnm{Chakresh Kumar} \sur{Singh}}

\author[2,3]{\fnm{Liubov} \sur{Tupikina}}

\author[4]{\fnm{Fabrice} \sur{L\'ecuyer}}

\author[5,6]{\fnm{Michele} \sur{Starnini}}

\author*[1,2]{\fnm{Marc} \sur{Santolini}}\email{marc.santolini@cri-paris.org}

\affil[1]{\orgname{Universit\'e Paris Cit\'e, Inserm, System Engineering and Evolution Dynamics}, \orgaddress{\postcode{F-75004}, \city{Paris}, \country{France}}}

\affil[2]{\orgname{Learning Planet Institute}, \orgaddress{\postcode{F-75004}, \city{Paris}, \country{France}}}

\affil[3]{\orgname{Nokia Bell Labs, France}}
\affil[4]{\orgname{Sorbonne Universit\'e, CNRS, LIP6}, \orgaddress{\city{F-75005 Paris}, \country{France}}}

\affil[5]{\orgname{CENTAI Institute}, \orgaddress{\city{Turin}, \country{Italy}}}

\affil[6]{\orgname{Departament de Fisica, Universitat Politecnica de Catalunya}, \orgaddress{\city{Campus Nord, 08034}, \country{Spain}}}



\abstract{From small steps to great leaps, metaphors of spatial mobility abound to describe discovery processes. Here, we ground these ideas in {formal} terms by systematically studying {scientific} knowledge mobility patterns. We use low-dimensional embedding techniques to create a knowledge space made up of 1.5 million articles from the fields of physics, computer science, and mathematics. By analyzing the publication histories of individual researchers, we discover patterns of knowledge mobility that closely resemble 
physical mobility. 
{In aggregate, the trajectories form mobility flows that} 
can be described by a gravity model, with jumps more likely to occur in areas of high density and less likely to occur over longer distances. 
We identify two types of researchers {from their individual mobility patterns}: interdisciplinary \textit{explorers} who pioneer new fields, 
and \textit{exploiters} who are more likely to stay within their specific areas of expertise. Our results suggest that spatial mobility analysis is a valuable tool for understanding knowledge evolution. 
}




\maketitle

\section*{Introduction}\label{sec:intro}

Quantifying the evolution of knowledge is crucial to understanding the past and predicting future innovations \cite{belikov2022prediction}, which ultimately lead to societal progress. At the forefront of scientific innovation are researchers recombining ideas to push the boundaries of the known \cite{iacopini_network_2018,ferreira2020quantifying}. 
With the exponential growth in the number of authors and publications \cite{bornmann2021growth,fortunato2018science}, novel methods are needed to represent and provide insights into knowledge development.

The increased access to large-scale publication datasets has provided opportunities to quantify the choices made by researchers and examine the factors governing the evolution of knowledge. 
By studying the citation patterns of researchers in their publications, studies have measured how conflicting ideas are pursued by researchers before they converge to a common consensus \cite{shwed_temporal_2010} or give way to new ideas \cite{lin2022new}. Other studies have focused on identifying `hot topics' in research \cite{liu2018hot}, quantifying knowledge flow patterns \cite{sun2020evolution} and memory effects in the evolution of knowledge \cite{yin_time_2017,pan2018memory}, or predicting the ultimate impact of a researcher \cite{wang2013quantifying,sinatra2016quantifying}. Similarly, keywords and phrases from publications can be leveraged to track the evolution of scientific ideas and fields \cite{chavalarias2013phylomemetic,battiston2019taking} or {quantify how scientists choose and shift their research focus over time} \cite{jia2017quantifying, zeng_increasing_2019}.
{For example, the Physics and Astronomy Classification Scheme (PACS) used 
in articles published by the American Physical Society can be exploited to 
study the ``essential tension''  between exploring the boundaries of a research area and exploiting previous work \cite{aleta_explore_2019}}. 
Finally, scientific credit among researchers and their mutual scientific interest (quantified by citations between papers and keywords, respectively) can be combined to improve the prediction of new scientific collaborations \cite{refId0}.

Therefore, studying the publication trajectories of researchers can help identify the multifaceted and complex processes underlying the evolution of knowledge. 
Such trajectories are often talked about {metaphorically}, for example when referring to some scientific advances as `great leaps' \cite{holden_federation_1974}.
{Here, we aim to explore the parallel between scientific and human mobility more formally, by leveraging insights from human mobility studies}. Using large-scale real-world data on human trajectories, previous studies have indeed uncovered several laws underlying human mobility. 
Despite heterogeneity in their movement, humans exhibit recurring patterns in their mobility \cite{wu_understanding_2021,ubaldi_heterogeneity_2021}. 
These patterns have been shown to give rise to scaling laws for the travel distance distribution \cite{barbosa_human_2018-1}. 
At the macroscopic level, the resulting flows between two locations follow a gravity model \cite{schlapfer2021universal}{, mimicking the Newtonian law of attraction between two masses at a given distance}.
Beyond jump distance, individuals show reproducible properties at the whole trajectory level. 
For example, individuals can be categorized into two classes, \textit{returners} and \textit{explorers}, depending on their propensity to come back to the same location or explore new ones \cite{pappalardo2015returners}. 
More generally, studies on both individual and collective mobility datasets have proposed various quantitative models explaining the dynamics of human mobility \cite{simini2021deep,alessandretti2020scales,barbosa_human_2018-1,schneider2013unravelling,simini2012universal,wilson_statistical_1967}. Crucially, these reproducible patterns are not unique to human mobility \cite{hills2015exploration}. 
Multiple studies across disciplines have found striking similarities between human mobility in geographic space and animals foraging \cite{hills2006animal}, insects swarms \cite{bonabeau1999swarm}, search methods in abstract environments such as memory space \cite{hills2012optimal}, organizational learning \cite{march_exploration_1991}, and cyberspace \cite{zhao2014scaling,hu2018life,barbosa2016returners}.

In the context of knowledge evolution, tools and data sources now abound for spatial representations. 
Natural language processing and embedding methods with metadata from publications such as citations, keywords, or abstracts, can be combined to exploit similarities between research publications and derive a low-dimensional representation of the scientific landscape. 
Such representations have been used to quantify the cognitive extent of ideas explored by researchers \cite{milojevic2015quantifying,milojevic_cognitive_2011} and the structural decomposition of journals \cite{peng2021neural}, giving insights into the structure of knowledge.

In this work, we solidify these intuitions into a quantitative framework to represent scientific knowledge and exploit metrics derived from human mobility to describe research trajectories. 
We use embedding methods on publication metadata from arXiv pre-prints to build a low-dimensional \textit{knowledge space} \cite{ying_modeling_2015}. 
We then track the mobility of disambiguated authors in this space using their publication records. 
We find that knowledge exploration shows striking similarities with human mobility in physical space. 
First, we show that scientific mobility in the knowledge space follows a gravity model, with jumps more
likely to occur in areas of high density and less likely to occur over longer distances. 
Second, we retrieve a dichotomy in knowledge exploration between {interdisciplinary} scientific \textit{explorers} -- more likely to {disrupt and} pioneer new fields -- and \textit{exploiters}, who tend to exploit {a particular area of expertise},
{mirroring} what is observed in spatial mobility between explorers and returners. 
{Finally, we discuss the usefulness of knowledge mobility analyses for the study of science and innovation, and discuss limitations and implications for future works.}

\section*{Results}\label{sec:results}


\subsection*{Scientific trajectories in the arXiv knowledge space}

{To build the knowledge space, we leverage} the arXiv dataset, encompassing $1,456,403$ scientific articles published online between 1992 and 2018 (see Methods, Figure~\ref{figSI:papers_year_field} and ref. \cite{singh2022quantifying}). 
Our interest in this dataset is two-fold. First, it has a clear and stable ontology for field tags, {which are used by authors to specify the relevant research area(s) covered by their articles}. 
There is a strong incentive for authors to document these tags as precisely as possible, in order for their article to appear in the right arXiv section searched by the target scientific community, {and in the relevant daily email digest that interested scientists can subscribe to}. 
Second, as a pre-print server, it has no editorial barrier or publication cost, creating a low threshold for publication. {This allows us to track the publication history of an author in a fine-grained manner, at the time they are considered finished, and irrespective of their perceived novelty}. 
As such, arXiv pre-prints can be thought of as tracking \textit{knowledge steps} to a high resolution, without requirements for novelty thresholds to be met.

We first build a spatial representation 
of the knowledge space formed by arXiv pre-prints.
The structure of this space is determined by the $175$ tags used by submitting authors to assign scientific sub-fields to articles. 
Articles can be assigned with one or more tags. For instance, an article can be tagged with Social and Information Networks (cs.si) and Physics and Society (physics.soc-ph). An article can thus be represented as a binary vector $X = (0,0,1,0,1,...,0)$ in the high dimensional $175$ 
sub-fields space, with $X_i =1$ if the article is assigned with the tag corresponding to scientific field $i$. 

\begin{figure}[tbp]%
	\centering
\includegraphics[width=0.95\textwidth]{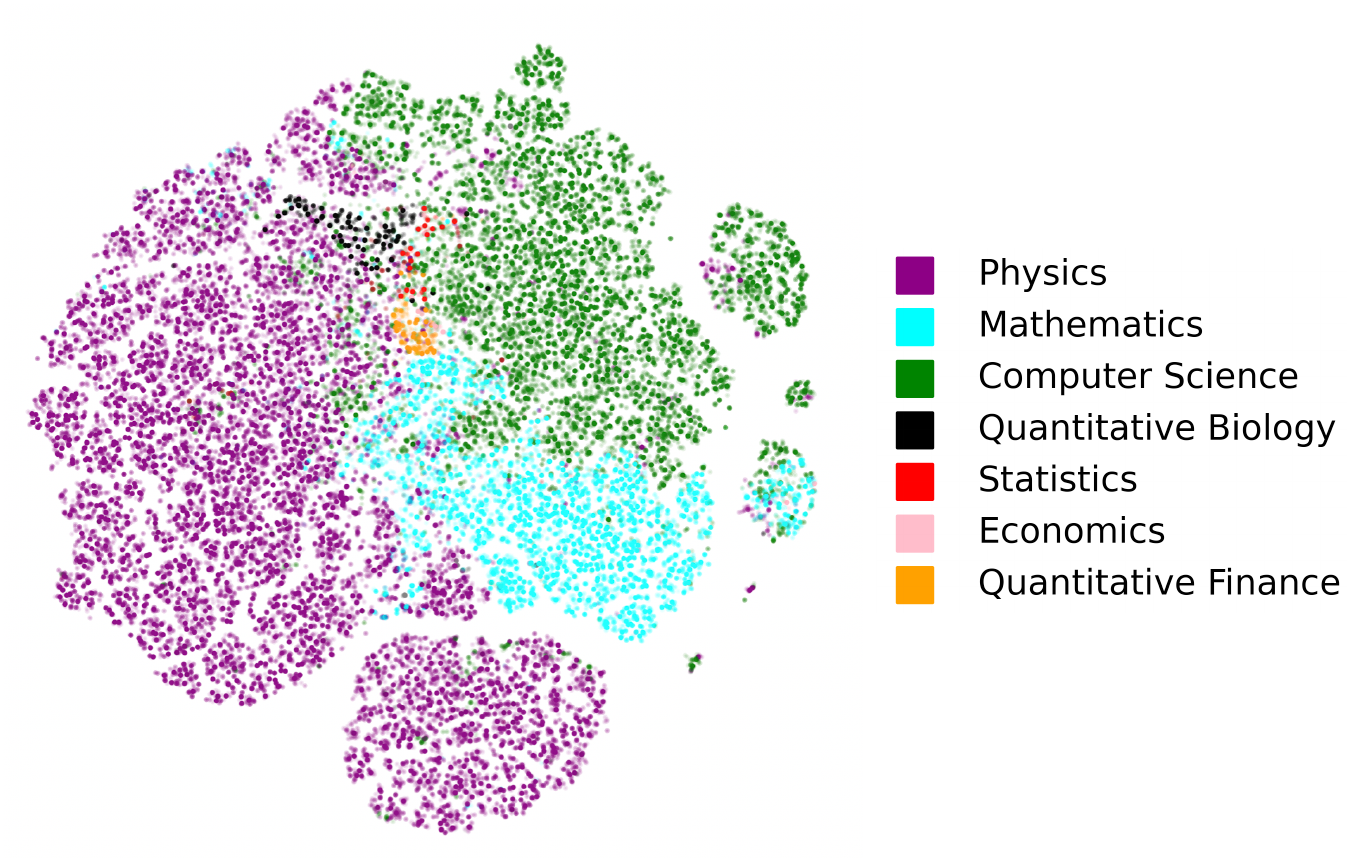}
	\caption{\textbf{Construction of the knowledge space.} 
We use the metadata from $1.45$ million articles posted on the arXiv, corresponding to the article field tags, authors, and timestamp. We build a high-dimensional $175$  space where each article is uniquely mapped through field tags corresponding to orthogonal dimensions. This high-dimensional space is finally embedded within a 2-dimensional {knowledge space} using the tSNE algorithm. 
Each point represents an article. Colors correspond to major academic fields in arXiv based on the first tag (i.e primary field) of the articles.
}
	\label{fig:knowledge_space}
\end{figure}

Since articles rarely combine more than a few tags (see Figure~\ref{figSI:sparsity}), 
the knowledge space is sparsely populated. Moreover, some tags co-occur frequently, creating redundant information \cite{singh2022quantifying}. Following these observations, we reduce the dimensionality of this initial space by embedding it into a low-dimensional space via the tSNE algorithm \cite{van2008visualizing,van2009learning} (see Methods). {In this study, we focus on a two-dimensional embedding to match traditional studies of human geographical mobility. In addition, we discuss the stability of the results with other embedding approaches in the Methods section}. Figure \ref{fig:knowledge_space} shows the resulting \textit{knowledge space}, where articles are represented as points colored according to their primary (first) field tag. We observe that articles belonging to the major fields from arXiv cluster into distinctive well-defined regions of the space, with interdisciplinary fields such as {Quantitative Biology (q-bio) or Quantitative Finance (q-fin)} located at the interface between related disciplines. 



\begin{figure}[tbp]%
	\centering
	\includegraphics[width=0.95\textwidth]{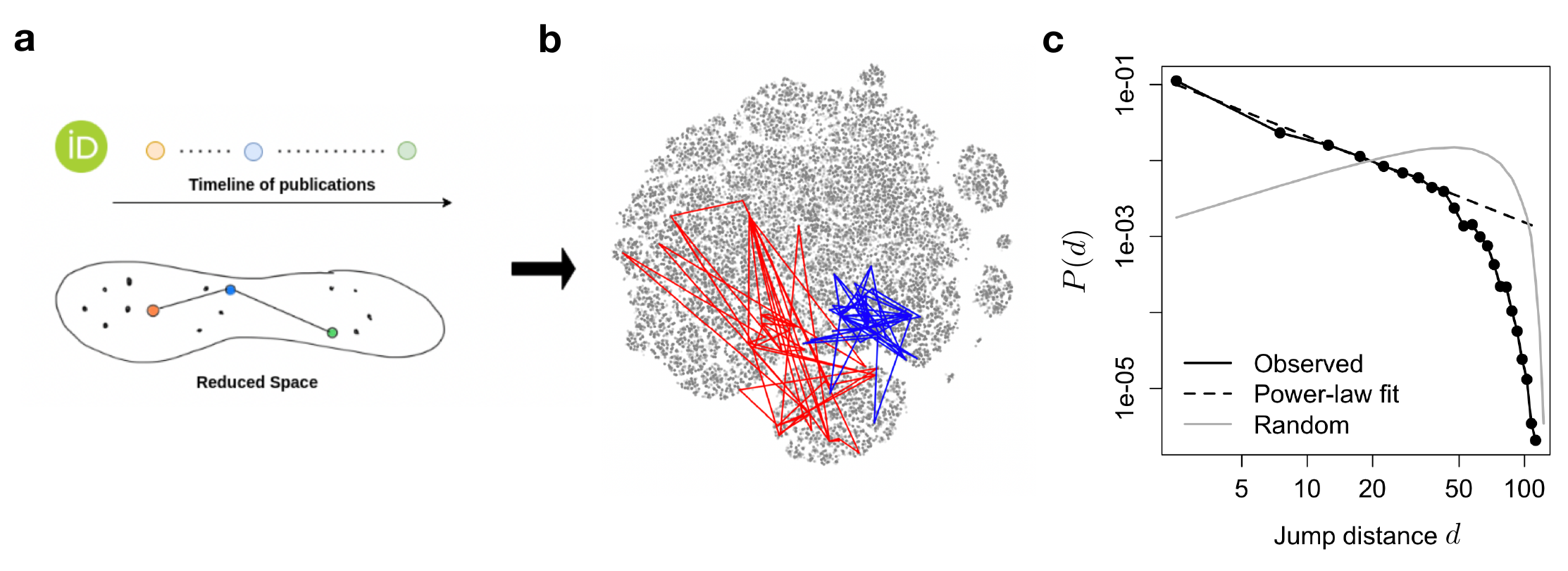}
	\caption{\textbf{Scientific mobility in the knowledge space.} \textbf{a.} The sequence of arXiv pre-prints of each researcher identifies a unique scientific trajectory in the knowledge space.  
 \textbf{b.} 
 We show two example trajectories indicating different behaviors of researchers. 
 \textbf{c.} Distribution of consecutive jump distances for authors  with at least 10 publications 
 ($N=11,826$). 
The dashed black line is a guide for the eye indicating a power-law behavior.
The gray line corresponds to the distribution of jump distances obtained when locations are selected at random across all possible visited locations, for each author.  
	}\label{fig:jump_distances}
\end{figure}

The chronological sequence of articles published by an author defines a sequence of locations in the knowledge space, tracing their \textit{scientific trajectory}  (Figure \ref{fig:jump_distances}a-b). 
In order to obtain high-quality trajectories, we 
select a sample of $11,826$ from a total of $50,402$ disambiguated researchers for which we have a unique ORCID identifier, and who published at least $10$ articles. 
Within a trajectory, two consecutive articles constitute a jump, with a length equal to the (euclidean) distance computed in the embedding, and duration equal to the number of days elapsed between the two articles. 
 If the authors were randomly jumping across all possible locations in the space, the jump distribution would follow a {bounded} distribution around a typical{, large} step size (Figure \ref{fig:jump_distances}c, gray line), according to a {pure} diffusive process. 
 Instead, Figure \ref{fig:jump_distances}c shows that the {jump distance} distribution is compatible with a power-law functional form, 
 with a cut-off {at large distances} due to the finite size of the space, differing significantly from a diffusive process.
 Importantly, we observe that this feature is robust with respect to different embedding techniques, see Figure \ref{figSI:tSNE_stability}. 
This indicates that, while the majority of jumps are small, {with researchers orbiting} relatively close to a particular research interest, 
a small fraction of jumps extend far into the knowledge space, standing for researchers crossing fields. 
In the next section, we investigate whether simple models of human mobility  can be compatible with the observed behavior. 



\subsection*{A Gravity Model of Scientific Mobility}


The observed fat-tail form (with a cutoff) of the jump size distribution is reminiscent of the inverse relation with distance observed in human mobility flows between two locations. 
This observation led to a simple and intuitive model in human mobility studies, the gravitation model, where the flux $F_{ij}$ between two locations $i$ and $j$ is proportional to the population sizes at $i$ and $j$ and inversely proportional to the distance $d_{ij}$ between them. 
{Earlier works on spatial distribution models and urban modelling \cite{WILSON1967253,senior1979gravity,wilson2013entropy} have shown that such a model can be functionally derived from statistical mechanics insights and empirical laws  
such as Zipf's law (derivation of gravity law is commented out)}
\cite{ribeiro_mathematical_2021}.  
When considering population-scale mobility in an origin-destination setting such as ours, the gravity model naturally emerges as the expectation of the distribution  
maximizing the entropy of mobility between two locations \cite{wilson_statistical_1967}.

\begin{figure}[tbp]%
	\centering
	\includegraphics[width=0.95\textwidth]{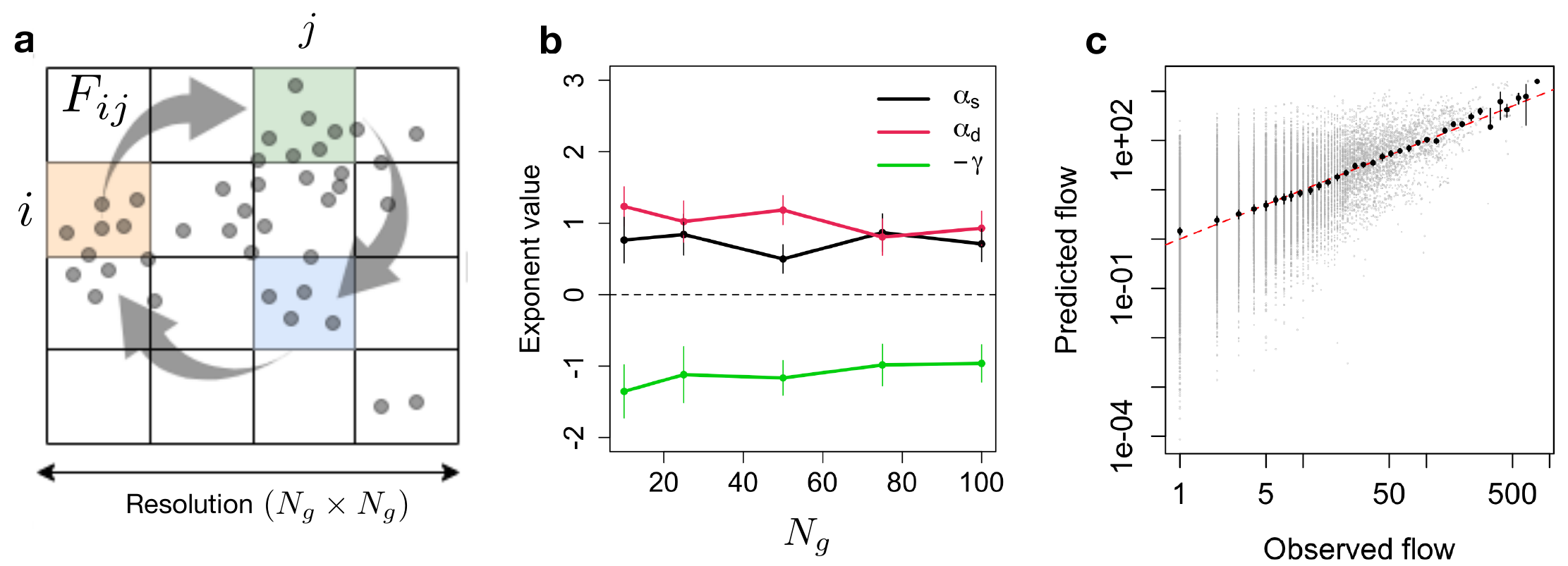}
	\caption{\textbf{A gravity model for scientific mobility.}
   \textbf{ a.} By dividing the projected space into a grid, we can calculate aggregated densities and study their effect on flow patterns between grid elements. The resolution of the grid is fixed by the number of cells along any dimension $N_g$. \textbf{b.} Fitted exponents from Eq. \eqref{eq:GM} for different grid resolution levels $N_g = 10$, $25$, $50$, $75$, $100$. \textbf{c.} Comparison between predicted and observed mobility flows at a grid size resolution $N_g=10$. }\label{fig:gravity_model}
\end{figure}

Much like the urban vs rural landscape, where populations conglomerate into a few, dense regions corresponding to urban areas, there are denser regions in the low-dimensional knowledge space, corresponding to more investigated areas. 
However, unlike cities and administrative areas, we do not have a clear definition of boundaries in the knowledge space. 
Here, we use a simple box/container model by defining a grid of size $N_g \times N_g$ covering the knowledge space, where $N_g$ is a parameter quantifying the resolution level, and {population counts are aggregated} at the grid level (Figure \ref{fig:gravity_model}a).
We then define a 
gravity model to predict the observed flow $F_{ij}$ between two grid locations $i$ and $j$ in the knowledge space, defined as the number of scientists jumping from grid location $i$ to location $j$, {by using a rolling time window of $5$ years}:

 \begin{equation}\label{eq:GM}
    \Tilde{F}_{ij} = G \frac{V_i^{\alpha_s}V_j^{\alpha_d}}{d_{ij}^{\gamma}},
\end{equation}

where $\Tilde{F}_{ij}$ is the predicted flow between locations $i$ and $j$, $G$ is a normalization constant, $d_{ij}$ is the distance between locations $i$ and $j$, and $V_i$ and $V_j$ (visits) are the numbers of authors who have published an article in locations $i$ and $j$ during the {$5$ previous years}.
The exponents $\alpha_s$,  $\alpha_d$, 
and $\gamma$ introduce non-linear scalings, such as crowding effects for the number of visits, where higher densities lead to sublinear ($\alpha<1$) or superlinear ($\alpha>1$) increase in flow. 

Figure \ref{fig:gravity_model}b shows the values of the exponents obtained by fitting Eq. \eqref{eq:GM} to the empirical flows $F_{ij}$ at different resolution levels $N_g$ (see Methods). Overall, we find a remarkable stability across grid sizes, with {coefficients} 
close to $1$. 
The quality of fit is shown in Figure \ref{fig:gravity_model}c, comparing predicted flows with observed flows, with a Pearson correlation value of $r=0.58$, indicating that the model explains $r^2=33.6\%$ of the variance of the mobility flows in the knowledge space. We note that the observed correlation is larger than the ones observed for real-world mobility (with $r$ between $0.03-0.49$, see \cite{simini2021deep}). {Finally, we find that beyond tSNE, the gravity model is able to represent flows of scientific mobility for different embedding techniques, with qualitatively similar exponents (see Figures \ref{figSI:gravity_fit_UMAP}, \ref{figSI:gravity_fit_PacMap})}.

\subsection*{Scientific explorers vs exploiters}




When jumping to their next article, researchers can move to a novel {region of the space}, or return back 
a previous one. 
That is, {in our framework} 
researchers choose between exploring a new scientific field or exploiting the previous knowledge they built. 
While such behaviors {can} lead to similar jump distribution patterns, they will impact more general statistics about the full trajectory, such as the extent of spatial territory covered. 
Previous studies have uncovered such a heterogeneity {between individual trajectories} in human mobility patterns, highlighting a dichotomy between returners, who gravitate around a small number of locations, and explorers, who rather move to new locations. {These results have been found to hold} both for spatial \cite{pappalardo2015returners}, as well as virtual \cite{barbosa_human_2018-1} contexts. 
{Here}, we explore whether such a heterogeneity exists in the context of knowledge exploration.
\begin{figure}[tbp]%
	\centering
	\includegraphics[width=\textwidth]{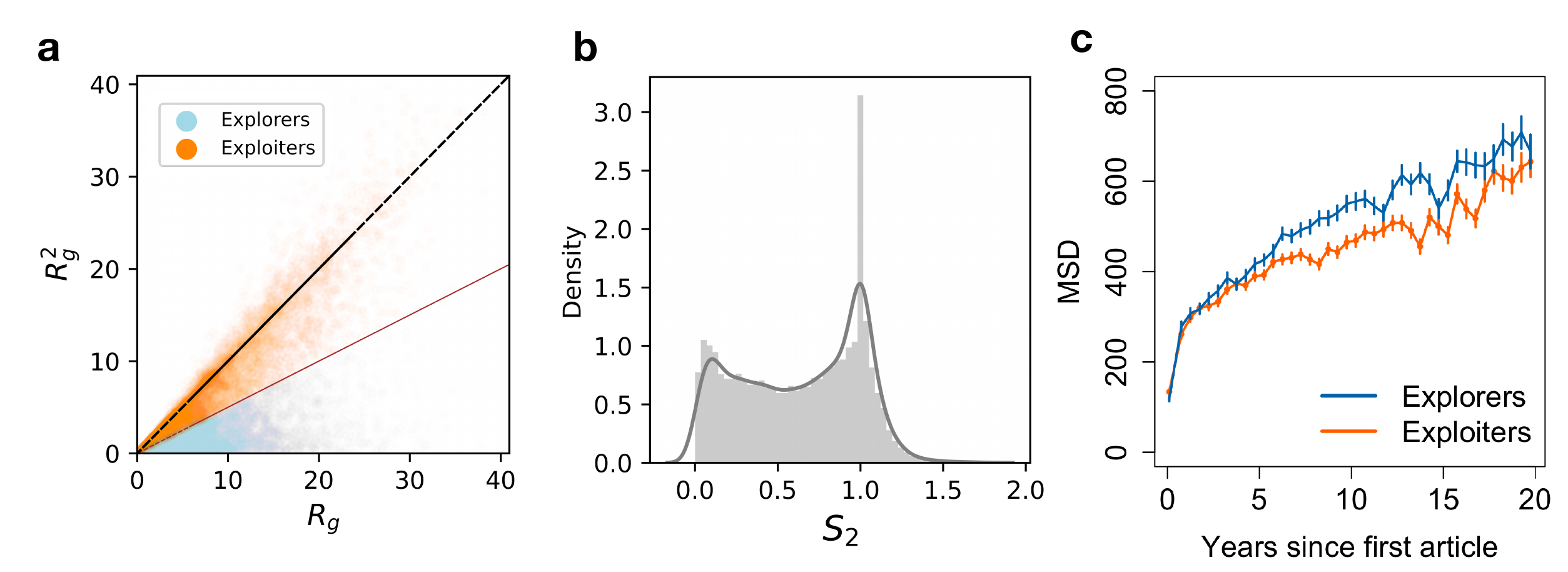}
	\caption{\textbf{Explorers and exploiters in the knowledge space.} \textbf{a.} Comparison of the radius of gyration $R_g^2$, with the center of mass computed from the 2 most visited locations, with the full radius of gyration $R_g$ (see Methods). We find a dichotomy between exploiters (orange) and explorers (blue), using the bisector method to classify them \cite{pappalardo2015returners}. \textbf{b.} Distribution of $S_2 = R_g^2 / R_g$, further highlighting the dichotomy as a bimodality of exploiters (close to 1) and explorers (close to 0). \textbf{c.} Comparison of the mean squared displacement as a function of time since their first article for explorers and exploiters.
	}\label{fig:retexp}
\end{figure}

{To} assess the extent of territory covered by a trajectory, {we} study the radius of gyration $R_g$, defined as the average distance of visited locations to their center of mass (see Eq. \eqref{eq_Rg} in Methods). 
By limiting to the top $k$ most visited locations, one can define the corresponding radius of gyration $R_g^k$ (Eq. \eqref{eq_Rgk}) and compare it with {the full} $R_g$ to {evaluate} the extent to which the trajectory returns to a few ($k$) locations. 
Figure \ref{fig:retexp}a shows the comparison of the total radius of gyration $R_g$ and $R_g^2$ across researchers. 
We find that researchers can be roughly grouped into two main classes: 
\textit{exploiters}, whose $R_g^2$ value is comparable to $R_g$ (points along the diagonal), and \textit{explorers} whose $R_g^2$ is considerably smaller than total $R_g$ (points closer to the x-axis). 
The two classes are
more evident when considering the distribution of $S_2 = R_g^2/R_g$, showing two peaks corresponding to the two populations of explorers and exploiters (Figure \ref{fig:retexp}b). {This bimodality disappears when considering larger values of $k$ (Figure \ref{figSI:retexp_k}), supporting the use of $k=2$ to distinguish the two classes.}

{By design,} the difference between explorers and exploiters will affect the research space spanned by scientific trajectories over time. {While the radius of gyration considers the gravitation of a researcher around a particular center of attraction (the center of gravity), other measures focus on the dynamics of departure from an original starting. 
In mobility analysis, this is typically quantified by the mean squared displacement (MSD) \cite{klafter2011first}, a quantity that tracks the average distance traveled from the starting location over time (see Methods). The particular interest in MSD stems from the fact that simple diffusion processes in homogeneous spaces observe a functional scaling with time, $\text{MSD}(t) \sim t^\beta$, with the exponent $\beta$ indicating a super- or sub-diffusive process. In our case,}
we find that, while both classes make jumps of similar size and duration (Figure~\ref{figSI:retexp_dists}) and have a similar sublinear MSD growth,
explorers span a larger fraction of the knowledge space early in their career, as indicated by a faster MSD growth between 5 and 15 years (Figure~\ref{fig:retexp}c). 
{This difference decreases in the later phase of their career (around 20 years),
indicating that researchers tend to explore mostly in the middle of their academic life}, while senior scientists tend to exploit more their previous research. {This finding comforts prior observations that scientists become less disruptive and more critical of emerging work as they age \cite{cui_aging_2022}}.

\begin{figure}[tbp]%
	\centering
	\includegraphics[width=0.7\textwidth]{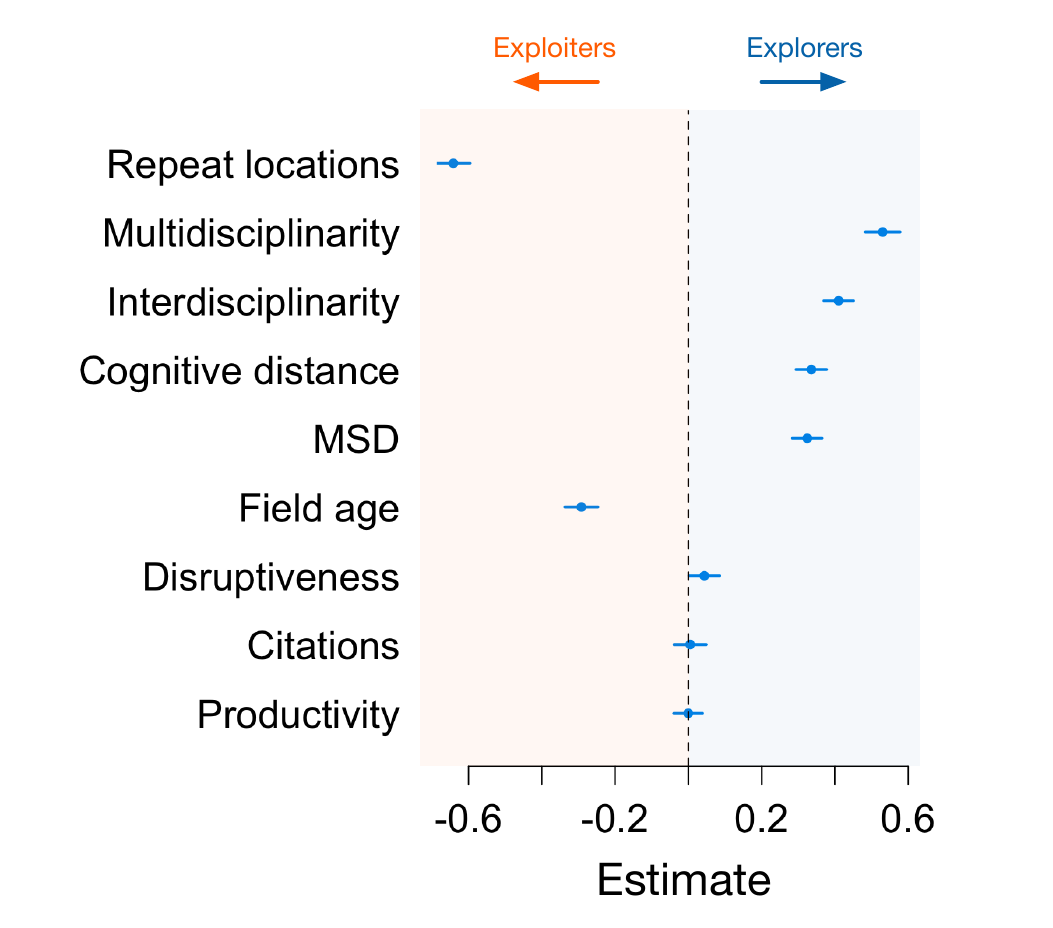}
	\caption{\textbf{Characteristics of explorers.} 
	We compute a logistic regression of a binary variable $y$ indicating whether an individual is an explorer ($y=1$) or an exploiter ($y=0$), for different characteristics of the researchers. For each attribute (further defined in Methods), we show the estimate and $95\%$ confidence interval of the standardized coefficient of the regression, controlling for the number of articles and the main field of interest of the researcher. 
    The differences between the two classes are all significant with p-values smaller than $10^{-5}$, except for disruptiveness ($p=0.04$), citations and productivity which are non-significant ($p>0.05$). 
    Repeat locations correspond to the proportion of jumps of size 0.  Multidisciplinarity is the total number of unique field tags used across articles, and interdisciplinarity is the average number of field tags per article. Cognitive distance is the maximum disciplinary distance spanned by the researcher, and Field age is the minimum normalized age of the fields across articles published by a researcher, both quantities being defined in \cite{singh2022quantifying}. MSD denotes the maximum mean square displacement achieved through the career of researchers. Disruptiveness is the maximum percentile of disruptive index achieved by {researchers} across their articles, when compared to the whole of arXiv. Citations correspond to the logarithm of the maximum number of citations received by the articles of the researcher. Finally productivity is the average yearly number of articles of the researcher.} 
	\label{fig:regression}
\end{figure}


{Lastly,} beyond differences in mobility patterns, we ask whether there are other characteristics that distinguish exploiters and explorers. 
To answer this question, we perform a logistic regression to predict if an individual {researcher} is an explorer as a function of several attributes. {To account for different trajectory lengths and field-specific behavior, we control for the {total} number of articles published and the {area} of interest (Figure \ref{fig:knowledge_space}) in which the author has published the most}. 
Figure~\ref{fig:regression} shows that, in line with the previous observation, explorers navigate broader regions of space
, as measured by their maximum MSD achieved throughout their career, while exploiters tend to remain at the same location
, measured by the proportion of their jumps being of distance $0$. 

Moreover, explorers cover more disciplines both within
and across 
articles, and these disciplines tend to be cognitively distant
, {i.e. they are far in the field tag co-occurrence network} \cite{singh2022quantifying}.
{When considering the association with specific developmental stages of scientific fields \cite{singh2022quantifying}, we find that explorers publish in the earlier stages of a field's evolution 
, a marker of pioneering activity and innovative work. Comforting this observation, we find a slightly higher disruptiveness for explorers ($p=0.04$), 
a quantitative marker of innovative works quantifying the extent to which articles citing an article of interest also cite its sources (low disruptiveness) or not (high disruptiveness) \cite{park_papers_2023}, measured here by the percentile of their most disruptive article}. 
Finally, we observe that explorers and exploiters show similar impact, measured by the maximum citations obtained in one of their articles, and yearly productivity. {We note however that results for the citation-based metrics are to be taken with care, as the citation network is only considering within-arXiv citations, and is therefore very incomplete \cite{clement_use_2019} and subject to field-specific habits}.



\section*{Discussion}\label{sec:discussion}


In this study, we show that methods from mobility analysis applied to a low-dimensional representation of a knowledge space 
can help understand the scientific mobility of researchers. 
Using data from 1.5M articles from the pre-print repository arXiv across 30 years, we find that the mobility patterns of researchers resemble those found in human mobility studies. 
Flows between {different} regions of the knowledge space follow a gravity model, with an inverse relation to distance. 
{This result is not an artefact from a particular representation, as it holds across various embedding parameters and methods (see Figures~\ref{figSI:tSNE_stability},\ref{figSI:gravity_fit_UMAP},\ref{figSI:gravity_fit_PacMap}). 
Furthermore, the model accuracy outperforms empirical results from human mobility studies \cite{simini2021deep}, showing that despite its simplicity, this model is a promising foundation to build on future work. 
In addition, by analyzing individual trajectories, we find that researchers can be categorized into \textit{exploiters}, whose trajectories are bound to a particular area of the knowledge space, and \textit{explorers}, who jump across boundaries and pioneer novel fields. 
This dichotomy is reminiscent of the ``essential tension'' between tradition and innovation in scientific research, where the desire to explore new promising areas is counterbalanced by the need to capitalize on the work done in the past \cite{kuhn_essential_1979,aleta_explore_2019}. Here we identify this tension by uncovering two types of knowledge mobility patterns through the bimodality observed in $R_g^2/R_g$.}  


{When considering the properties of scientific trajectories in the knowledge space}, we observe that the mobility patterns of explorers and exploiters show sub-diffusive regimes. 
{Theoretically}, when considering the mobility of an individual in a homogeneous space, such as the initial hypercube or a regular lattice, the MSD follows a linear regime if the second moment of the step size distribution and the first moment of the waiting-time probability distribution are finite \cite{klafter2011first}. In our case, the observed deviation from a linear MSD (Figure \ref{fig:retexp}) may be due to the heavy-tailed waiting time probability distribution of the two groups of researchers \cite{klafter2011first} (Figure \ref{figSI:retexp_dists}). 
Another possibility is that the multi-scale nature of the knowledge space, as a complex and evolving cognitive construct, may be responsible for this {trapped-like} behavior. Further investigation is needed to determine which of these approaches is more suitable for explaining the observed non-linearity in MSD behavior.

We assumed stable categories of exploiters and explorers using the full trajectory of researchers, yet there can be variation throughout their career. For example, we observed aging patterns within trajectories, with MSD  
of explorers and exploiters showing similar values within the first 5 years after their first publication, after which MSD values for explorers are significantly larger (Figure~\ref{fig:retexp}c). 
This could indicate that explorers go through two phases: a first phase where they are staying within a few most visited locations, followed by an exploratory behaviour towards other locations. Such a behavior could be formally captured using the concept of ``intermittent behavior'' from stochastic processes \cite{lanoiselee_unraveling_2017}. Future work could investigate such temporal patterns across research trajectories, for example by using time windows or the convex hull method to analyse dynamic profiles at a finer scale, and assess whether phases might correspond to institutional constraints, with some environment fostering the individual development towards more exploratory patterns.


Our framework relies on the method used to define the knowledge space.
There is no ground truth in the use of embedding methods, and each can bias results towards specific idiosyncratic properties. 
However, we have shown that a variety of parametric (tSNE) and non-parametric (UMAP, PaCMap) methods yield qualitatively similar results, both in terms of long-tailed jump distribution (Figure~\ref{figSI:tSNE_stability}) and gravity model fit (Figs~\ref{figSI:gravity_fit_UMAP}-\ref{figSI:gravity_fit_PacMap}). 
This indicates that, despite some variations coming from the structure of the space itself, the general mobility patterns uncovered here are not space representation artifacts.

Our work is focused on a dataset of arXiv pre-prints. 
{This dataset provides} a high precision for the identification of subfields, which is useful for both the construction of the knowledge space and the computation of features such as field age \cite{singh2022quantifying}. {Yet, it is limited in overall size, fields covered, and incompleteness of the citation network. 
It is therefore yet unclear how our findings generalize to other disciplines, for example when considering the humanities or social sciences.} 
Future work should explore the {reproducibility of our findings across larger and diverse datasets, leveraging} other field identification methods, such as the ones using Natural Language Processing {\cite{beltagy_scibert_2019}}.

While our study is focused on the description of individual trajectories, most articles are team-authored \cite{wuchty2007increasing}, and chaperoning patterns are fundamental to scientific careers \cite{sekara2018chaperone}. 
Therefore, future studies could study the couplings between individual trajectories, leading to correlated patterns and ultimately collective {flows}. 
In addition, the gravity model could be extended to incorporate variables corresponding to local attributes, such as impact (e.g. through citations), field age, devoted funding, etc. These features might act as biasing forces shaping collective flows towards certain areas of the knowledge space. On a macroscopic level, these fields can affect mobility, in the same way that force fields affect the trajectories of particles in physics. Novel methods based on deep learning, such as a Deep Gravity Model \cite{simini2021deep}, {coupled with more extensive data on citations and funding}, could help extend our work. 

Overall, the insights gained from leveraging a mobility analysis in the knowledge space could help study the effect of policies on knowledge exploration and exploitation, with applications for funding agencies and more generally the evaluation of research.

\section*{Methods}
\label{sec:methods}

\subsection*{Overview of the arXiv dataset}

In our study we use a previously published dataset consisting of article metadata from the arXiv
preprint repository \cite{singh2022quantifying}. 
The dataset consists of $1,456,403$ articles published between $1992$ and $2018$, covering mainly the fields of physics, mathematics and computer science, and to a lesser extent Quantitative Biology, Statistics, Finance, Economy, and Engineering. 
We note in particular the important rise of Computer Science articles in the past decade, with the number of articles published bound to soon outweigh those from the physics field, which was at the core of the early arXiv usage (Figure~\ref{figSI:papers_year_field}).

When {uploading} an article, the submitting author selects a main, primary tag identifying the core discipline, along with secondary tags if needed. {In most cases, arXiv require authors who are submitting papers to a subject category for the first time to get an endorsement from an established arXiv author, as a quality control mechanism}. The tags span $175$ predefined subfields, such as Quantum Algebra (math.QA) or Signal Processing (eess.SP), all indicated on the website's main page. These subfields have remained relatively stable in time \cite{singh2022quantifying}. Moreover, there is a strong incentive for authors to select appropriate fields, as arXiv proposes a subscription service to a daily digest email system to automatically receive novel submitted articles {containing a specific field tag}. As such, the tag system is directly tied to a relevant audience for the publishing individual, incentivizing for an accurate self-report. 


\subsection*{Low dimensional embeddings}

To reduce the dimensionality of the initial $175$-dimensional field space, we use the tSNE algorithm, an unsupervised, parametric dimensionality reduction technique that retains the local data structure in the latent space \cite{van2008visualizing,van2009learning}. 
The tSNE method captures much of the local structure of the high-dimensional data, while also revealing global structure such as the presence of clusters at different scales. The visualisation of the resulting embedding of the arXiv knowledge space into a two 
dimensional space is shown in Figure~\ref{fig:knowledge_space}. Each point corresponds to one of the $49,575$ observed combinations of field tags within arXiv articles. We note that permutations of tags map to the same point, so that our analysis does not depend on the order of tags.


For the implementation of the tSNE algorithm we use the \texttt{scikit-learn} package in Python \cite{pedregosa_scikit-learn_2011}. The dimension of the embedded space is set to $2$.
The main parameters of the embedding method, such as learning rate, number of iterations and early exaggeration parameters are set to default values.
In order to test the robustness of the tSNE embedding to varying parameters, we generated tSNE mapping for different perplexity levels $p$ (signifying the nearest neighbors) and learning rate parameters $LR$ of the algorithm, and plotted the pairwise distance distribution between randomly sampled points across different settings (Figure~\ref{figSI:tSNE_space}). We find remarkable stability across various parameters of the tSNE suggested in \cite{wang2021understanding}, including perplexity levels, indicating that choosing different tSNE parameters would not strongly affect the results.


\subsection*{Robustness with respect to the embedding method}

In order to assess the robustness of our results, we tested the impact of different embedding methods, parameters, as well as subsamples of the data on the jump distribution (Figure~\ref{figSI:tSNE_stability}). Beyond tSNE, we evaluated the robustness of our analysis using PaCMAP and UMAP embeddings. 
 
 The Uniform Manifold Approximation and Projection (UMAP) \cite{mcinnes2018umap} has its theoretical foundations in manifold theory and topological data analysis. At a macroscopic level, UMAP uses local manifold approximations and fuzzy simplicial sets to construct topological representations of data in high and low dimensions. It then minimizes the cross-entropy between the two topological representations to find an optimal lower-dimensional representation. UMAP can also be understood as a k-neighbour based graph learning algorithm that finds the best representation of weighted graphs in lower dimensions.

 The Pairwise Controlled Manifold Approximation (PaCMAP) algorithm \cite{wang2021understanding} is also a graph-based technique that identifies three sets of pairs namely - neighbor pairs, mid-near pairs and further pairs. It then systematically optimizes its loss function using a custom gradient descent algorithm to find a lower dimensional representation that preserves both local and global structures.

 We find that the jump distance distributions shows a similar long-tail decay for both methods (Figure~\ref{figSI:tSNE_stability}a). 
 In addition, we find that the gravity model has stable results in UMAP (Figure~\ref{figSI:gravity_fit_UMAP}) and PaCMAP (Figure~\ref{figSI:gravity_fit_PacMap}) contexts, though we find an overall smaller exponent for the distance, closer to $\gamma \simeq 0.5$.\\

\subsection*{Fitting procedure for the gravity model}

In order to fit the gravity model, we used a linear regression of log-transformed variables. For each resolution level, we first computed {for each year starting in $1997$ the number of jumps between a source cell $i$ and a target cell $j$ (with $i \neq j$) and number of articles published in each cell in the $5$ preceding years}. In order to account for low sample size, we used a pseudo-count of $1$ added to raw visit counts: $V_i \leftarrow V_i + 1$. We then computed the natural logarithm of all quantities, and used these log-transformed values for the regression analysis. Since there is a much larger number of small flow values compare with large flow values (Figure~\ref{fig:gravity_model}), we used a binning technique to avoid overfitting our model to low flow values. For this, we cut the obtained log-flow values into $100$ bins containing an equal number of points, and merged bins with the same breakpoints, resulting in $42$ final bins. We then averaged the log-transformed features (flow, visits, distance) within these bins, and used these average values to fit the gravity model, using the \texttt{lm} function from R 4.2.2. Residuals of the model are shown in Figure~\ref{figSI:gravity_fit} and are normally distributed.

\subsection*{Radius of gyration}

The radius of gyration measures the typical size of the territory spanned by the trajectory of an individual. To compute it,
 we first define the center of mass $R_{cm}$ of the trajectory across locations $i=0,1, \ldots, n$: 
\begin{equation}\label{eq_Rcm}
R_{cm} = \frac{\sum_{i=1}^{n} M_i R_i}{\sum_{i=1}^{n}  M_i},
\end{equation}

where 
$M_i$ is the frequency of visitation of each location $i$, i.e. the number of times location $i$ is visited by the individual, and $R_i$ is {the radius vector characterising the location in the knowledge space with respect to the chosen center of coordinates}.  
The radius of gyration is then defined as the characteristic distance from the center of mass:

    
\begin{equation}\label{eq_Rg}
R_g = \sqrt{\frac{\sum_{i=1}^n M_i(R_i-R_{cm})^2}{\sum_{i=1}^n  M_i}}.   
\end{equation}

In order to estimate the influence of a few locations over the trajectory, we define the $k$-th radius of gyration by considering only the top $k$ most visited location:

\begin{equation}\label{eq_Rgk}
R_g^k = \sqrt{\frac{\sum_{i=1}^{k} M_i(R_i-R_{cm}^k)^2}{\sum_{i=1}^{k}  M_i}},
\end{equation}
where $R_{cm}^k$ is the center of mass using the top $k$ most visited locations. 


\subsection*{Mean squared displacement }

The mean squared displacement (MSD) at time $t$ for a trajectory {is defined as the deviation of the position of a walker (in our case, a researcher) with respect to a reference position over time}:
\begin{equation}\label{eq_MSD}
\text{MSD} (t) = \langle \vert x(t) -x(0) \vert^2 \rangle 
\end{equation}
where 
$x(t)$ stands for the position of researcher at time $t$ since the first article
, and $x(0)$ stands for the starting point of the trajectory. 

\subsection*{Logistic regression for explorers vs. exploiters}

In order to explore the characteristics associated with explorers in Figure~\ref{fig:regression}, we compute a logistic regression with dependent variable $y_i$, a binary variable indicating whether an individual $i$ is an explorer ($y_i = 1$) or an exploiter ($y_i = 0$), and independent variables various individual features $x_i$. 
We control for the main field $F_i$ in which the author has published (given by the most represented field tag across their articles), as well as the number of articles $N_i$ of the researcher. The fields were encoded as factors. We used the \texttt{glm} function in R to fit the model \texttt{$y_i \sim x_i + N_i + F_i$}, with parameter \texttt{family = binomial} set to a logistic regression. Regression summaries were obtained using the \texttt{summ} function from the \texttt{jtools} package in R, with parameters \texttt{scale=T} to standardize the regression coefficients by scaling and mean-centering input data, and \texttt{confint=T} to obtain $95\%$ confidence intervals.

\subsection*{Innovation, disruptiveness and impact}

To measure the innovative level of a work, we used two methods. First, we computed for each article how early it occurs within the fields that it mentions. To do so, we computed the minimum \textit{rescaled time} (RT) across its associated field tags, using the method described in \cite{singh2022quantifying}. {The rescaled time is a normalized quantity that allows us to associate an article to a developmental stage of a field (early, peak, or late phase) even when fields have drastically different rise and fall durations}. We then computed for each researcher the minimum RT value achieved across their articles, defining the ``Field age''.

Second, we used another independent method to assess the innovative potential of the articles. This method leverages how disruptive an article is by comparing the attention it receives compared to the articles it cites. Citation data was obtained from \cite{clement_use_2019}. The disruptive index (DI) was then computed using the method from \cite{wu_large_2019} for each article. For each author, we computed the maximum DI across their articles. Finally, we computed the percentile of the obtained value across articles to compute the disruptiveness of an author.

Finally, for each author $i$, we computed the maximum number of citations $\hat{c}_{i}$ received by any of their articles across their career. Since citation counts are distributed with a heavy-tailed function, we used the transformation $\log(\hat{c}_{i}+1)$ to quantify the impact.

\subsection*{Cognitive Distance}

We observe in Figure \ref{fig:regression} that compared to exploiters, explorers use a larger number of field tags per article, as well as a larger number of unique tags across their articles. However some tags might be more closely related than others in terms of research area, which the simple measure for linear estimate of tags used does not differentiate. 
To account for this effect, we use the network based cognitive distance measure from \cite{singh2022quantifying}, where the cognitive distance $C_{ij}$ between field tags $i$ and $j$ is the weighted distance along the shortest path between tags $i$ and $j$ in the tag co-occurrence network. 

\section*{Acknowledgements}

We thank Emma Barme for early discussions on this topic, as well as Ariel Lindner and Robert Ward for advice on relevant literature. Thanks to the Bettencourt Schueller Foundation long term partnership, this work was partly supported by the CRI Research Fellowship to Marc Santolini.

\newpage

\renewcommand\thefigure{S\arabic{figure}}    
\setcounter{figure}{0}

\newpage

\section*{Supplementary Figures}

\begin{figure}[h]%
	\centering
 	\includegraphics[width=0.8\textwidth]{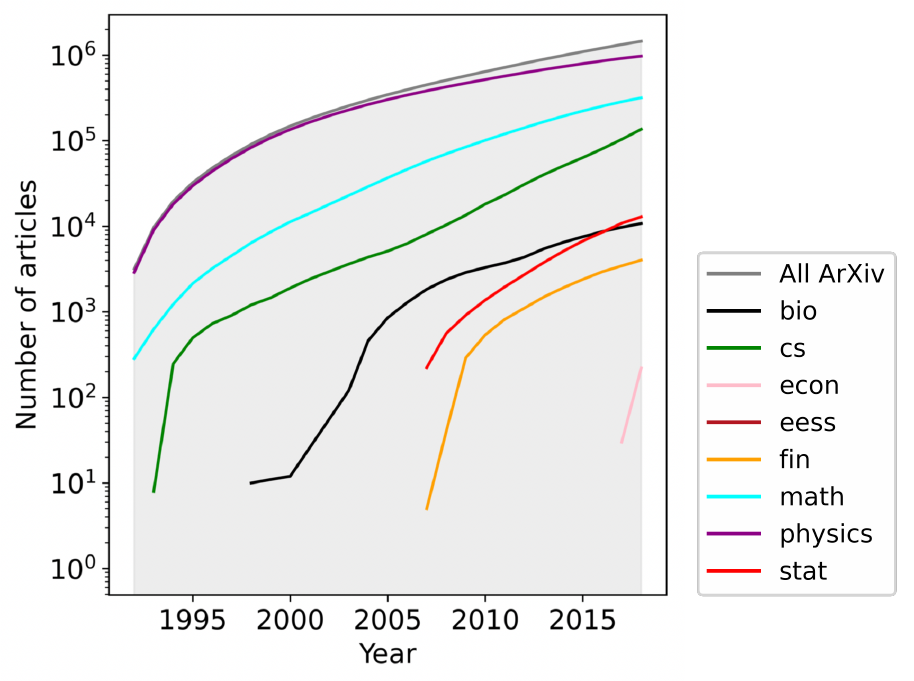}
	\caption{\textbf{{Growth in the number of articles submitted to the major fields of arXiv} between 1992 and 2018}. We show the {cumulative} number of articles over time for the $9$ major fields in arXiv, totalling $1456403$ articles in total. {Fields consist of: Quantitative Biology (bio), Computer Science (cs), Economy (econ), Electrical Engineering and Systems Science (eess), Quantitative Finance (fin), Mathematics (math), Physics (physics), and Statistics (stat)}. We use a log-scaling for an easier visualisation of counts for the more recent fields.  Beyond the early dominance of Physics, within which the arXiv usage originally emerged, we see the rapid{, exponential} rise of Computer Science in the early 2,000s. 
 }\label{figSI:papers_year_field}
\end{figure}

\begin{figure}[h]%
	\centering
	\includegraphics[width=0.5\textwidth]{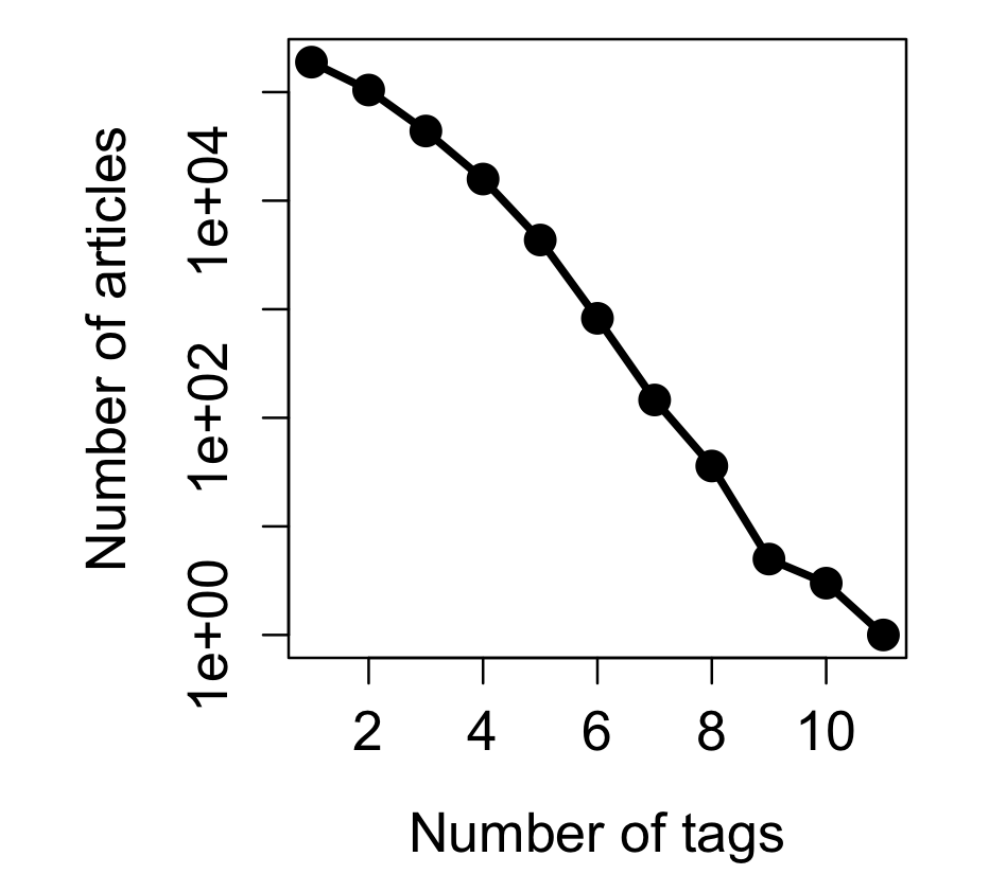}
	\caption{\textbf{Sparsity of the high dimensional space.} Number of articles in arXiv with a given number of tags. Most papers have $10$ or less tags, which is much less than the dimensionality of the space ($N=175$), meaning that most possible locations in the space are not populated.}\label{figSI:sparsity}
\end{figure}

\begin{figure}[h]%
	\centering
	\includegraphics[width=0.6\textwidth]{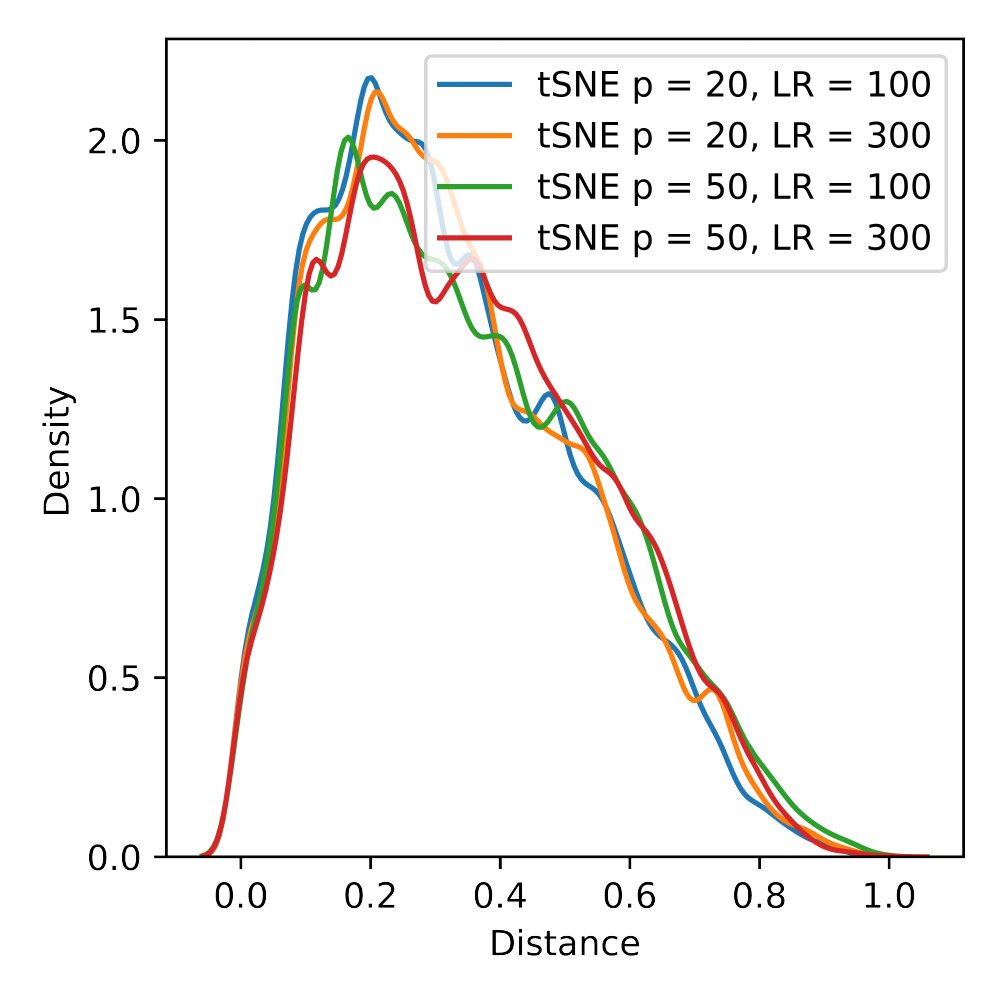}
	\caption{\textbf{Comparison of pairwise distance distributions across various tSNE parameters.} Density distributions of the distance between randomly sampled pairs of points in different tSNE embeddings, with parameters shown in legend.} 
 \label{figSI:tSNE_space}
\end{figure}

\begin{figure}[h]%
	\centering
	\includegraphics[width=0.9\textwidth]{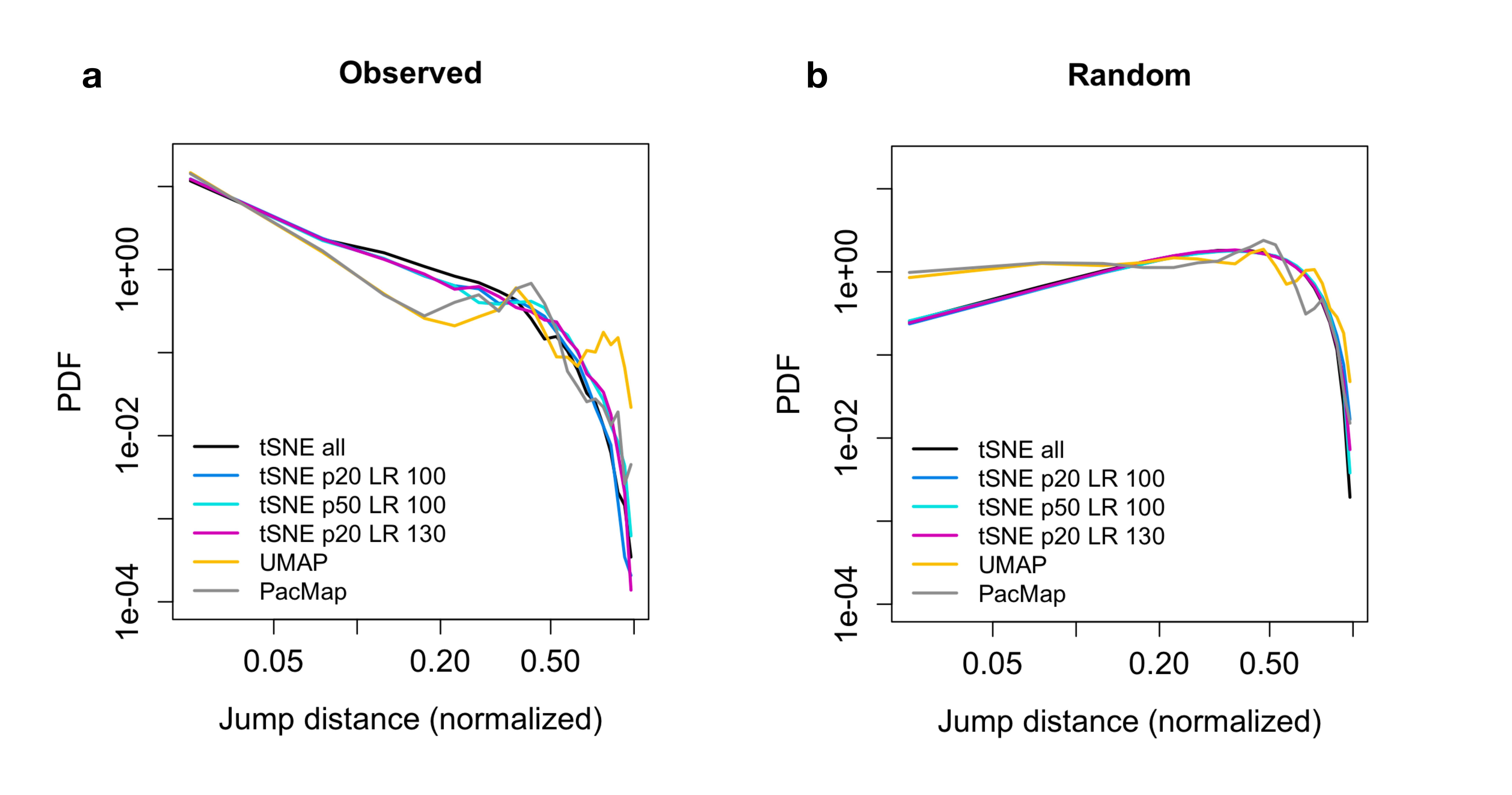}
	\caption{\textbf{Comparison of jump distributions across embeddings methods.} \textbf{a.} We show the variation of the jump distribution when using different parameters for the tSNE embedding ($p$ is perplexity, and $LR$ is Learning Rate) as well as other embedding methods (PaCMap and UMAP). {In order to compare between the different embeddings, the Jump distance is normalized by the maximum distance for each embedding}. \textbf{b.} For each author, we select random locations across all accessible points in the embedding (i.e. unique existing locations in the dataset) and plot the corresponding randomized jump distributions across embeddings.} 
 \label{figSI:tSNE_stability}
\end{figure}




\begin{figure}[h]%
	\centering
	\includegraphics[width=0.95\textwidth]{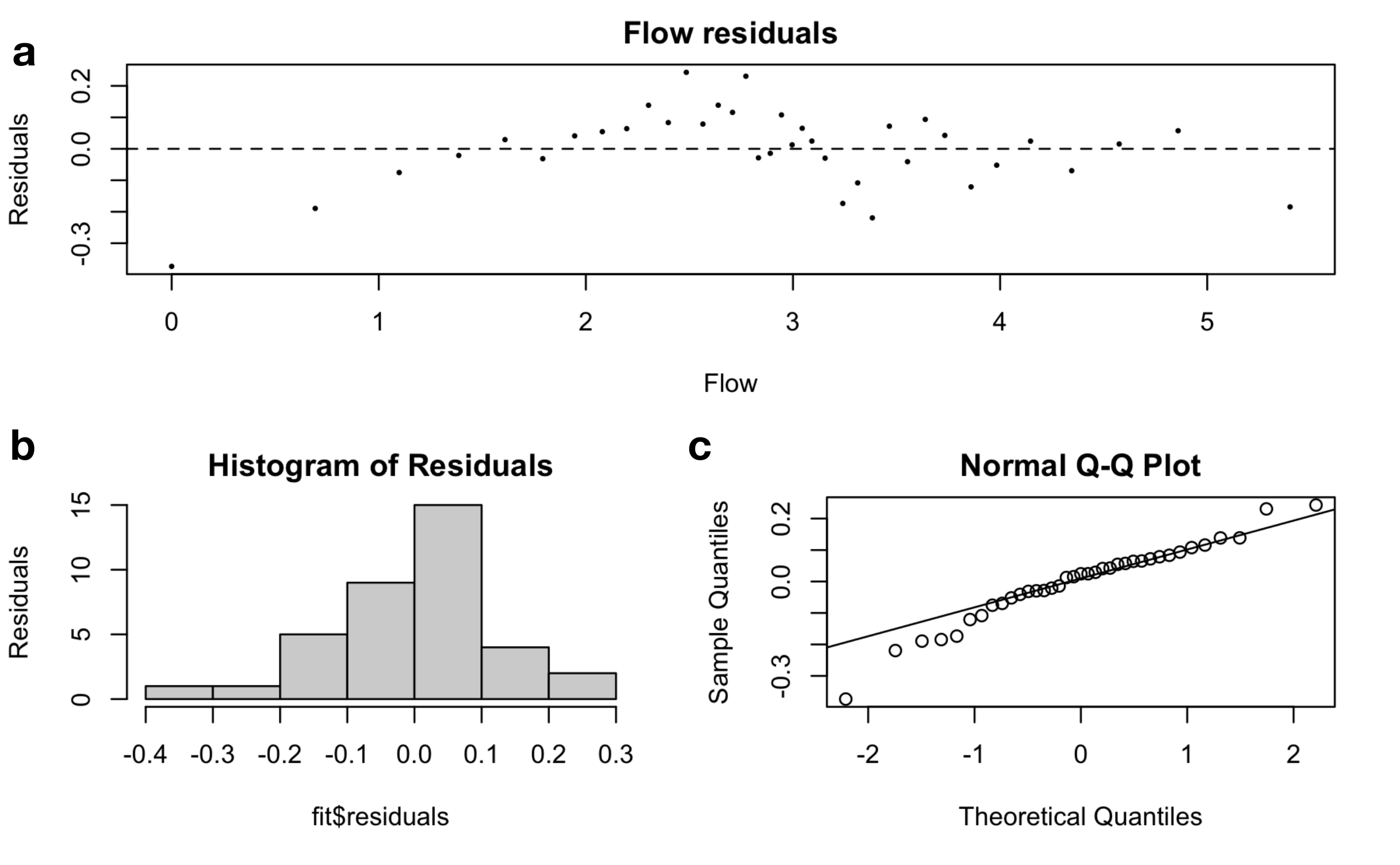}
	\caption{\textbf{Residual plots for the gravity model.} {We test that the assumption of normality of residuals hold in our regression analysis. \textbf{a} Value of residuals as a function of the fitted (log) flow values. We find that very low flow values are slightly over-estimated, which might stem from the pseudo-counting method used. \textbf{b} Histogram of the residuals. \textbf{c} The residuals are normally distributed, as can be assessed using a Q-Q plot.}}\label{figSI:gravity_fit}
\end{figure}

\begin{figure}[h]%
	\centering
	\includegraphics[width=0.95\textwidth]{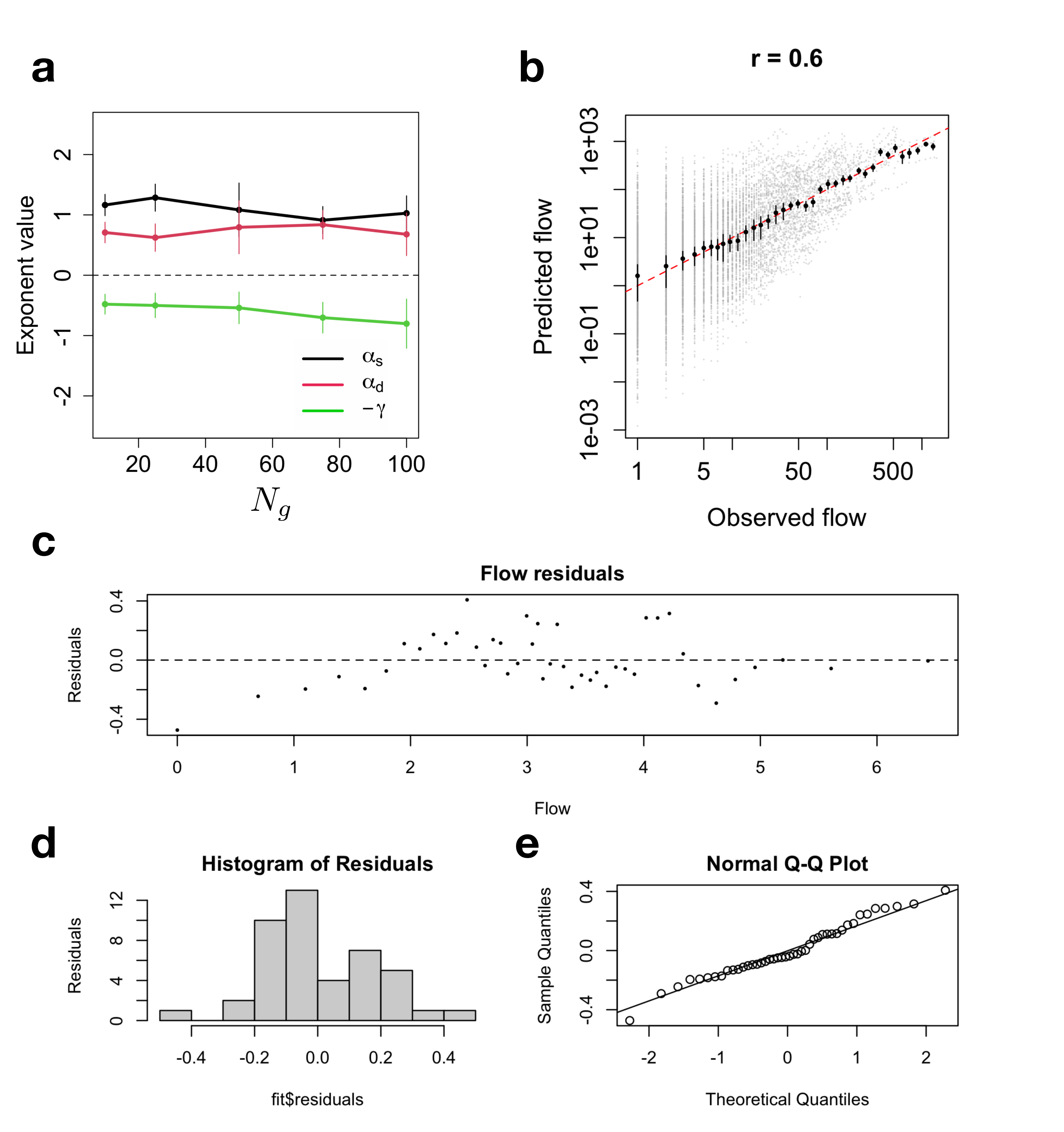}
	\caption{\textbf{Results of the fit of the gravity model for the UMAP embedding.} \textbf{a-b.} Same than Figure \ref{fig:gravity_model}b-c. \textbf{c-e.} Same as Figure \ref{figSI:gravity_fit}. }\label{figSI:gravity_fit_UMAP}
\end{figure}

\begin{figure}[h]%
	\centering
	\includegraphics[width=0.95\textwidth]{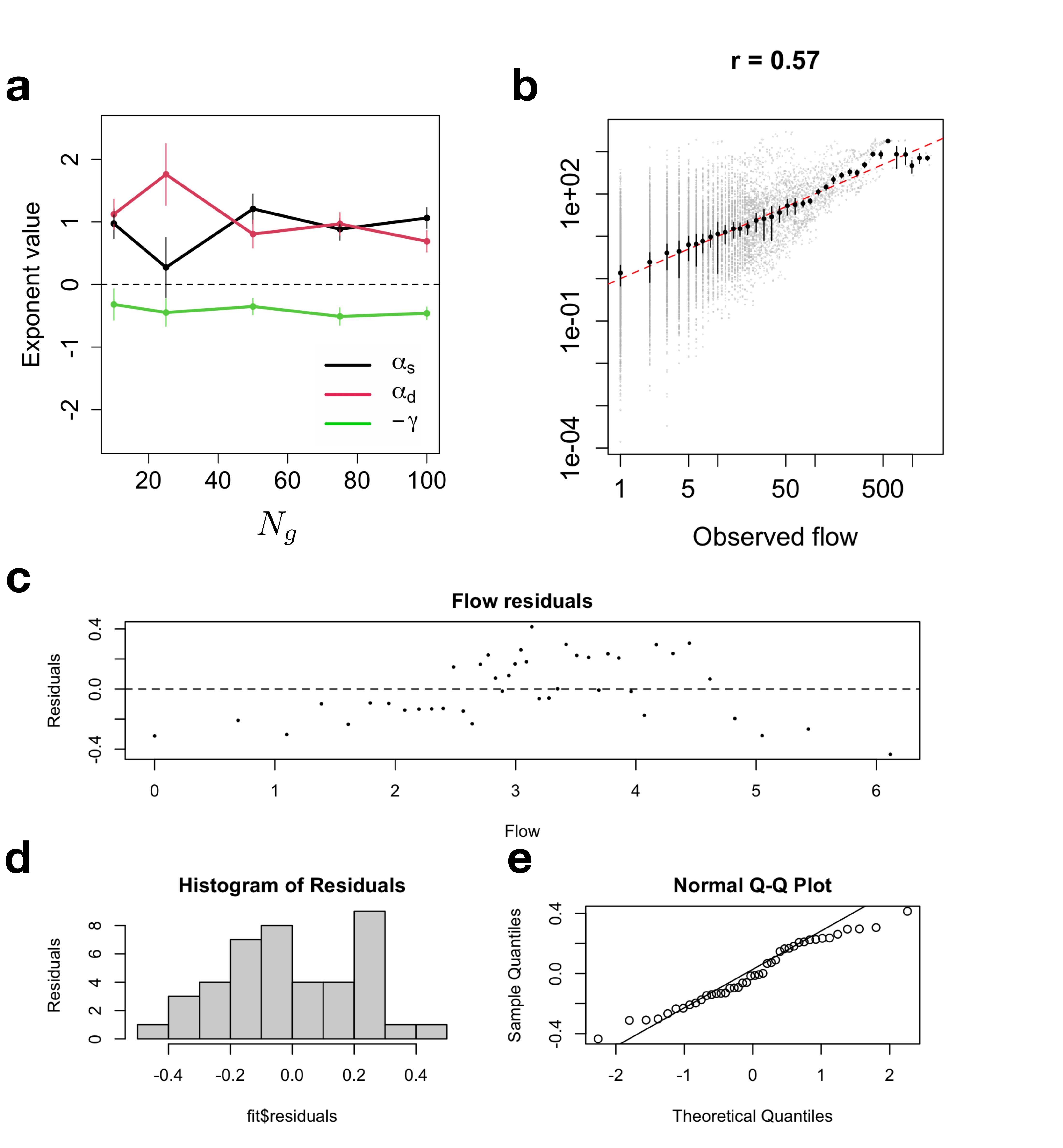}
	\caption{\textbf{Results of the fit of the gravity model for the PaCMAP embedding.} \textbf{a-b.} Same than Figure \ref{fig:gravity_model}b-c. \textbf{c-e.} Same as Figure \ref{figSI:gravity_fit}.   }\label{figSI:gravity_fit_PacMap}
\end{figure}


\begin{figure}[h]%
	\centering
	\includegraphics[width=\textwidth]{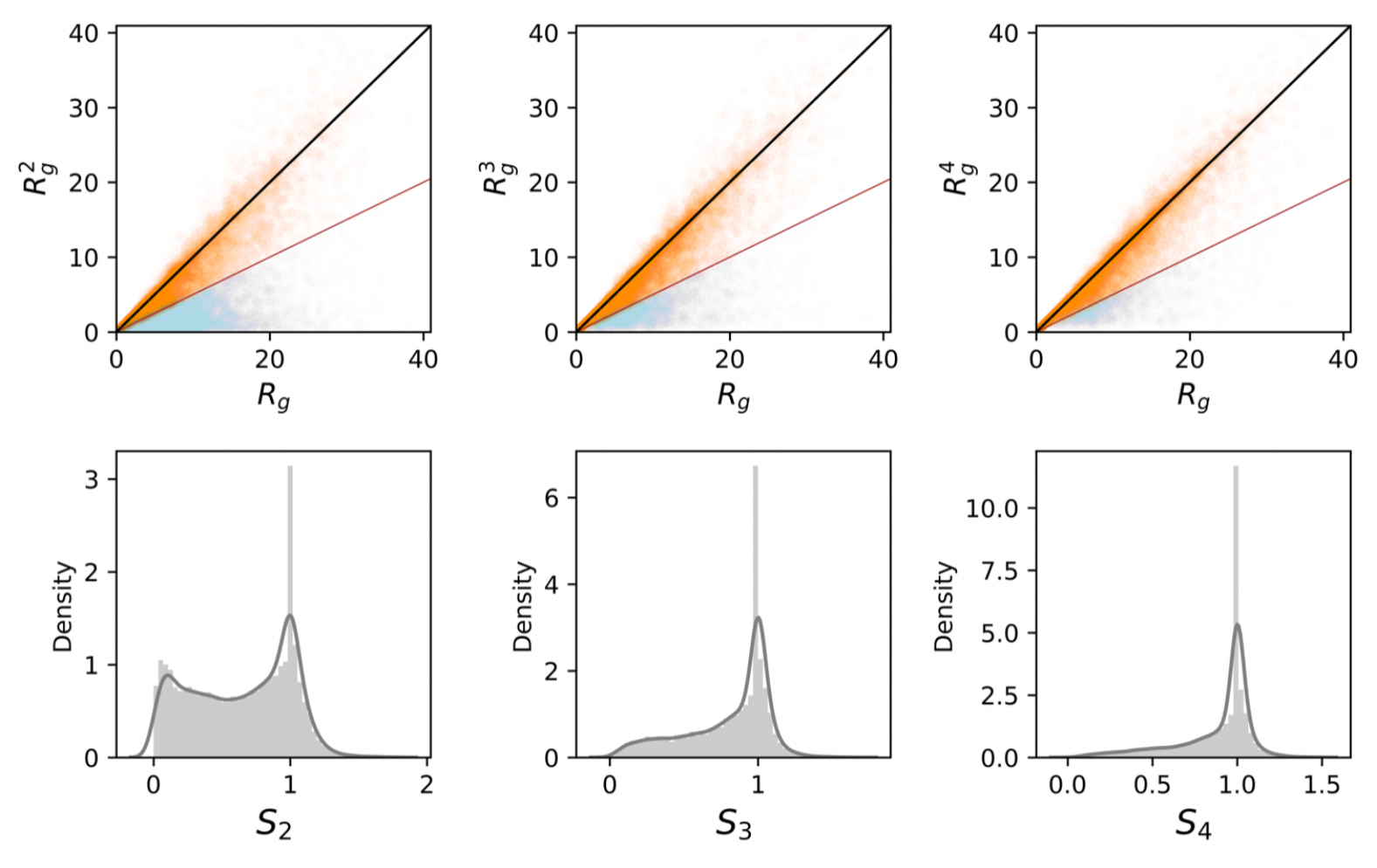}
	\caption{\textbf{{Explorers and exploiters for larger values of $k$.}}
{ Top row: same as Figure \ref{fig:retexp}a, for $k=2,3,4$. The red line corresponds to $R_g^k$ - $R_{g}/2 = 0$, with points above the line (orange) corresponding to exploiters, and points below the line (blue) corresponding to explorers. Bottom row: same as Figure \ref{fig:retexp}b, for $k=2,3,4$. The bimodality observed for $k=2$ vanishes for larger values of $k$.}}\label{figSI:retexp_k}
\end{figure}




\begin{figure}[h]%
	\centering
	\includegraphics[width=\textwidth]{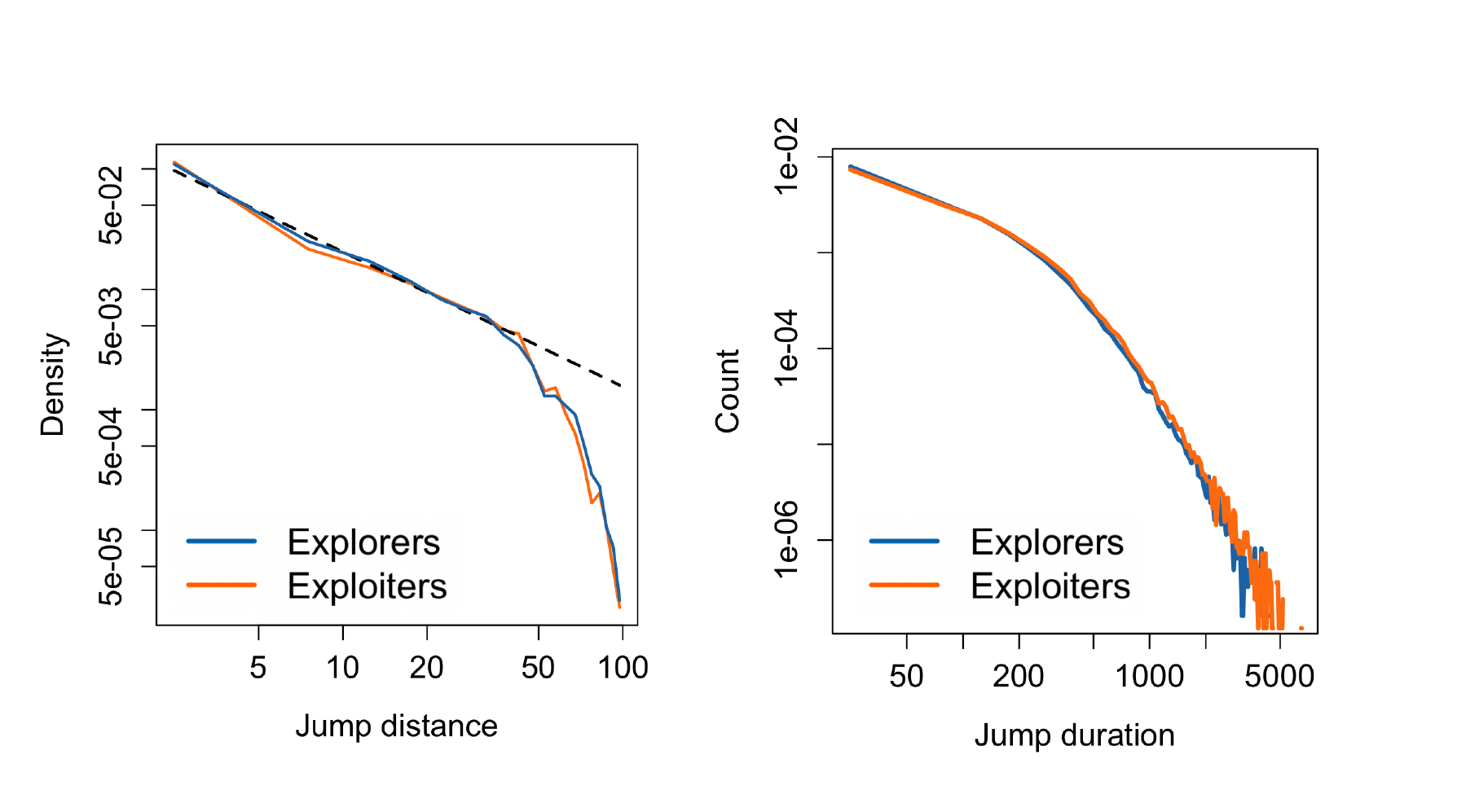}
	\caption{\textbf{Comparing jump distributions between explorers and exploiters.} The jump distance (left) and jump duration (right, in days) distributions are similar between explorers and exploiters (log-log plots).}\label{figSI:retexp_dists}
\end{figure}

\begin{figure}[h]%
	\centering
	\includegraphics[width=\textwidth]{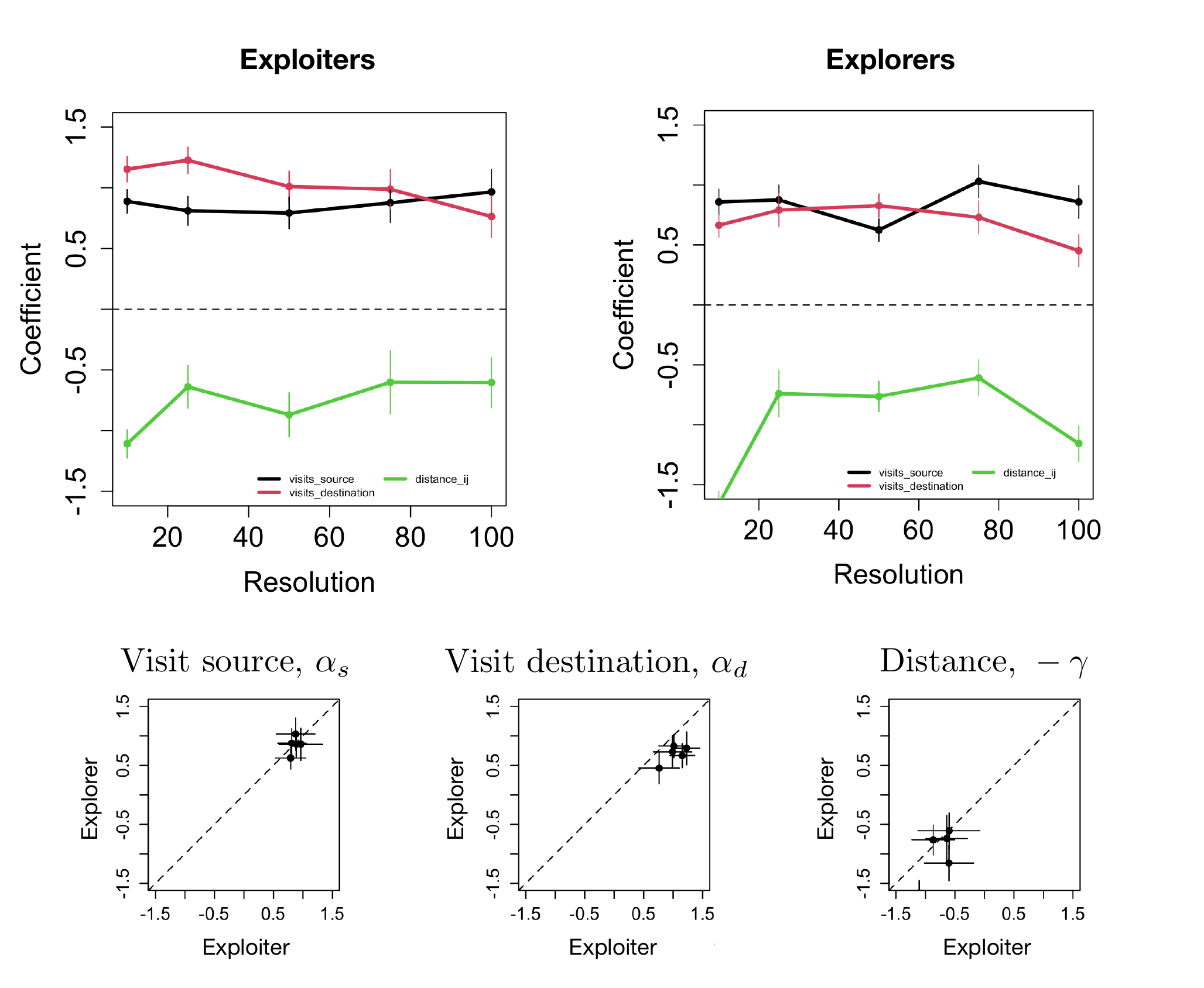}
	\caption{\textbf{Gravity model for explorers and exploiters.} We show the results of the gravity model fitting procedure applied to exploiters only (top left) and explorers only (top right). We then compare the obtained exponents for visit source, visit destination and distance across resolutions between both categories (bottom). Error bars denote standard error from the regression. }\label{figSI:retexp_gravity}
\end{figure}

\clearpage
\bibliography{ref}


\begin{thebibliography}{62}
\ifx \bisbn   \undefined \def \bisbn  #1{ISBN #1}\fi
\ifx \binits  \undefined \def \binits#1{#1}\fi
\ifx \bauthor  \undefined \def \bauthor#1{#1}\fi
\ifx \batitle  \undefined \def \batitle#1{#1}\fi
\ifx \bjtitle  \undefined \def \bjtitle#1{#1}\fi
\ifx \bvolume  \undefined \def \bvolume#1{\textbf{#1}}\fi
\ifx \byear  \undefined \def \byear#1{#1}\fi
\ifx \bissue  \undefined \def \bissue#1{#1}\fi
\ifx \bfpage  \undefined \def \bfpage#1{#1}\fi
\ifx \blpage  \undefined \def \blpage #1{#1}\fi
\ifx \burl  \undefined \def \burl#1{\textsf{#1}}\fi
\ifx \doiurl  \undefined \def \doiurl#1{\url{https://doi.org/#1}}\fi
\ifx \betal  \undefined \def \betal{\textit{et al.}}\fi
\ifx \binstitute  \undefined \def \binstitute#1{#1}\fi
\ifx \binstitutionaled  \undefined \def \binstitutionaled#1{#1}\fi
\ifx \bctitle  \undefined \def \bctitle#1{#1}\fi
\ifx \beditor  \undefined \def \beditor#1{#1}\fi
\ifx \bpublisher  \undefined \def \bpublisher#1{#1}\fi
\ifx \bbtitle  \undefined \def \bbtitle#1{#1}\fi
\ifx \bedition  \undefined \def \bedition#1{#1}\fi
\ifx \bseriesno  \undefined \def \bseriesno#1{#1}\fi
\ifx \blocation  \undefined \def \blocation#1{#1}\fi
\ifx \bsertitle  \undefined \def \bsertitle#1{#1}\fi
\ifx \bsnm \undefined \def \bsnm#1{#1}\fi
\ifx \bsuffix \undefined \def \bsuffix#1{#1}\fi
\ifx \bparticle \undefined \def \bparticle#1{#1}\fi
\ifx \barticle \undefined \def \barticle#1{#1}\fi
\bibcommenthead
\ifx \bconfdate \undefined \def \bconfdate #1{#1}\fi
\ifx \botherref \undefined \def \botherref #1{#1}\fi
\ifx \url \undefined \def \url#1{\textsf{#1}}\fi
\ifx \bchapter \undefined \def \bchapter#1{#1}\fi
\ifx \bbook \undefined \def \bbook#1{#1}\fi
\ifx \bcomment \undefined \def \bcomment#1{#1}\fi
\ifx \oauthor \undefined \def \oauthor#1{#1}\fi
\ifx \citeauthoryear \undefined \def \citeauthoryear#1{#1}\fi
\ifx \endbibitem  \undefined \def \endbibitem {}\fi
\ifx \bconflocation  \undefined \def \bconflocation#1{#1}\fi
\ifx \arxivurl  \undefined \def \arxivurl#1{\textsf{#1}}\fi
\csname PreBibitemsHook\endcsname

\bibitem{belikov2022prediction}
\begin{barticle}
\bauthor{\bsnm{Belikov}, \binits{A.V.}},
\bauthor{\bsnm{Rzhetsky}, \binits{A.}},
\bauthor{\bsnm{Evans}, \binits{J.}}:
\batitle{Prediction of robust scientific facts from literature}.
\bjtitle{Nature Machine Intelligence}
\bvolume{4}(\bissue{5}),
\bfpage{445}--\blpage{454}
(\byear{2022})
\end{barticle}
\endbibitem

\bibitem{iacopini_network_2018}
\begin{barticle}
\bauthor{\bsnm{Iacopini}, \binits{I.}},
\bauthor{\bsnm{Milojević}, \binits{S.}},
\bauthor{\bsnm{Latora}, \binits{V.}}:
\batitle{Network dynamics of innovation processes}.
\bjtitle{Phys. Rev. Lett.}
\bvolume{120}(\bissue{4}),
\bfpage{048301}
(\byear{2018}).
\doiurl{10.1103/PhysRevLett.120.048301}.
\bcomment{arXiv:1707.04239 [physics]}
\end{barticle}
\endbibitem

\bibitem{ferreira2020quantifying}
\begin{bbook}
\bauthor{\bsnm{Ferreira}, \binits{M.R.}},
\bauthor{\bsnm{Reisz}, \binits{N.}},
\bauthor{\bsnm{Schueller}, \binits{W.}},
\bauthor{\bsnm{Servedio}, \binits{V.D.P.}},
\bauthor{\bsnm{Thurner}, \binits{S.}},
\bauthor{\bsnm{Loreto}, \binits{V.}}:
In: \beditor{\bsnm{La~Porta}, \binits{C.A.}},
\beditor{\bsnm{Zapperi}, \binits{S.}},
\beditor{\bsnm{Pilotti}, \binits{L.}} (eds.)
\bbtitle{Quantifying Exaptation in Scientific Evolution},
pp. \bfpage{55}--\blpage{68}.
\bpublisher{Springer},
\blocation{Cham}
(\byear{2020}).
\doiurl{10.1007/978-3-030-45784-6_5}
\end{bbook}
\endbibitem

\bibitem{bornmann2021growth}
\begin{barticle}
\bauthor{\bsnm{Bornmann}, \binits{L.}},
\bauthor{\bsnm{Haunschild}, \binits{R.}},
\bauthor{\bsnm{Mutz}, \binits{R.}}:
\batitle{Growth rates of modern science: a latent piecewise growth curve
  approach to model publication numbers from established and new literature
  databases}.
\bjtitle{Humanities and Social Sciences Communications}
\bvolume{8}(\bissue{1}),
\bfpage{1}--\blpage{15}
(\byear{2021})
\end{barticle}
\endbibitem

\bibitem{fortunato2018science}
\begin{barticle}
\bauthor{\bsnm{Fortunato}, \binits{S.}},
\bauthor{\bsnm{Bergstrom}, \binits{C.T.}},
\bauthor{\bsnm{B{\"o}rner}, \binits{K.}},
\bauthor{\bsnm{Evans}, \binits{J.A.}},
\bauthor{\bsnm{Helbing}, \binits{D.}},
\bauthor{\bsnm{Milojevi{\'c}}, \binits{S.}},
\bauthor{\bsnm{Petersen}, \binits{A.M.}},
\bauthor{\bsnm{Radicchi}, \binits{F.}},
\bauthor{\bsnm{Sinatra}, \binits{R.}},
\bauthor{\bsnm{Uzzi}, \binits{B.}}, \betal:
\batitle{Science of science}.
\bjtitle{Science}
\bvolume{359}(\bissue{6379}),
\bfpage{0185}
(\byear{2018})
\end{barticle}
\endbibitem

\bibitem{shwed_temporal_2010}
\begin{barticle}
\bauthor{\bsnm{Shwed}, \binits{U.}},
\bauthor{\bsnm{Bearman}, \binits{P.S.}}:
\batitle{The {Temporal} {Structure} of {Scientific} {Consensus} {Formation}}.
\bjtitle{Am Sociol Rev}
\bvolume{75}(\bissue{6}),
\bfpage{817}--\blpage{840}
(\byear{2010}).
\doiurl{10.1177/0003122410388488}.
\bcomment{Publisher: SAGE Publications Inc}
\end{barticle}
\endbibitem

\bibitem{lin2022new}
\begin{barticle}
\bauthor{\bsnm{Lin}, \binits{Y.}},
\bauthor{\bsnm{Evans}, \binits{J.A.}},
\bauthor{\bsnm{Wu}, \binits{L.}}:
\batitle{New directions in science emerge from disconnection and discord}.
\bjtitle{Journal of Informetrics}
\bvolume{16}(\bissue{1}),
\bfpage{101234}
(\byear{2022})
\end{barticle}
\endbibitem

\bibitem{liu2018hot}
\begin{barticle}
\bauthor{\bsnm{Liu}, \binits{L.}},
\bauthor{\bsnm{Wang}, \binits{Y.}},
\bauthor{\bsnm{Sinatra}, \binits{R.}},
\bauthor{\bsnm{Giles}, \binits{C.L.}},
\bauthor{\bsnm{Song}, \binits{C.}},
\bauthor{\bsnm{Wang}, \binits{D.}}:
\batitle{Hot streaks in artistic, cultural, and scientific careers}.
\bjtitle{Nature}
\bvolume{559}(\bissue{7714}),
\bfpage{396}--\blpage{399}
(\byear{2018})
\end{barticle}
\endbibitem

\bibitem{sun2020evolution}
\begin{barticle}
\bauthor{\bsnm{Sun}, \binits{Y.}},
\bauthor{\bsnm{Latora}, \binits{V.}}:
\batitle{The evolution of knowledge within and across fields in modern
  physics}.
\bjtitle{Scientific reports}
\bvolume{10}(\bissue{1}),
\bfpage{1}--\blpage{9}
(\byear{2020})
\end{barticle}
\endbibitem

\bibitem{yin_time_2017}
\begin{barticle}
\bauthor{\bsnm{Yin}, \binits{Y.}},
\bauthor{\bsnm{Wang}, \binits{D.}}:
\batitle{The time dimension of science: {Connecting} the past to the future}.
\bjtitle{Journal of Informetrics}
\bvolume{11}(\bissue{2}),
\bfpage{608}--\blpage{621}
(\byear{2017}).
\doiurl{10.1016/j.joi.2017.04.002}
\end{barticle}
\endbibitem

\bibitem{pan2018memory}
\begin{barticle}
\bauthor{\bsnm{Pan}, \binits{R.K.}},
\bauthor{\bsnm{Petersen}, \binits{A.M.}},
\bauthor{\bsnm{Pammolli}, \binits{F.}},
\bauthor{\bsnm{Fortunato}, \binits{S.}}:
\batitle{The memory of science: Inflation, myopia, and the knowledge network}.
\bjtitle{Journal of Informetrics}
\bvolume{12}(\bissue{3}),
\bfpage{656}--\blpage{678}
(\byear{2018})
\end{barticle}
\endbibitem

\bibitem{wang2013quantifying}
\begin{barticle}
\bauthor{\bsnm{Wang}, \binits{D.}},
\bauthor{\bsnm{Song}, \binits{C.}},
\bauthor{\bsnm{Barab{\'a}si}, \binits{A.-L.}}:
\batitle{Quantifying long-term scientific impact}.
\bjtitle{Science}
\bvolume{342}(\bissue{6154}),
\bfpage{127}--\blpage{132}
(\byear{2013})
\end{barticle}
\endbibitem

\bibitem{sinatra2016quantifying}
\begin{barticle}
\bauthor{\bsnm{Sinatra}, \binits{R.}},
\bauthor{\bsnm{Wang}, \binits{D.}},
\bauthor{\bsnm{Deville}, \binits{P.}},
\bauthor{\bsnm{Song}, \binits{C.}},
\bauthor{\bsnm{Barab{\'a}si}, \binits{A.-L.}}:
\batitle{Quantifying the evolution of individual scientific impact}.
\bjtitle{Science}
\bvolume{354}(\bissue{6312}),
\bfpage{5239}
(\byear{2016})
\end{barticle}
\endbibitem

\bibitem{chavalarias2013phylomemetic}
\begin{barticle}
\bauthor{\bsnm{Chavalarias}, \binits{D.}},
\bauthor{\bsnm{Cointet}, \binits{J.-P.}}:
\batitle{Phylomemetic patterns in science evolution—the rise and fall of
  scientific fields}.
\bjtitle{PloS one}
\bvolume{8}(\bissue{2}),
\bfpage{54847}
(\byear{2013})
\end{barticle}
\endbibitem

\bibitem{battiston2019taking}
\begin{barticle}
\bauthor{\bsnm{Battiston}, \binits{F.}},
\bauthor{\bsnm{Musciotto}, \binits{F.}},
\bauthor{\bsnm{Wang}, \binits{D.}},
\bauthor{\bsnm{Barab{\'a}si}, \binits{A.-L.}},
\bauthor{\bsnm{Szell}, \binits{M.}},
\bauthor{\bsnm{Sinatra}, \binits{R.}}:
\batitle{Taking census of physics}.
\bjtitle{Nature Reviews Physics}
\bvolume{1}(\bissue{1}),
\bfpage{89}--\blpage{97}
(\byear{2019})
\end{barticle}
\endbibitem

\bibitem{jia2017quantifying}
\begin{barticle}
\bauthor{\bsnm{Jia}, \binits{T.}},
\bauthor{\bsnm{Wang}, \binits{D.}},
\bauthor{\bsnm{Szymanski}, \binits{B.K.}}:
\batitle{Quantifying patterns of research-interest evolution}.
\bjtitle{Nature Human Behaviour}
\bvolume{1}(\bissue{4}),
\bfpage{1}--\blpage{7}
(\byear{2017})
\end{barticle}
\endbibitem

\bibitem{zeng_increasing_2019}
\begin{barticle}
\bauthor{\bsnm{Zeng}, \binits{A.}},
\bauthor{\bsnm{Shen}, \binits{Z.}},
\bauthor{\bsnm{Zhou}, \binits{J.}},
\bauthor{\bsnm{Fan}, \binits{Y.}},
\bauthor{\bsnm{Di}, \binits{Z.}},
\bauthor{\bsnm{Wang}, \binits{Y.}},
\bauthor{\bsnm{Stanley}, \binits{H.E.}},
\bauthor{\bsnm{Havlin}, \binits{S.}}:
\batitle{Increasing trend of scientists to switch between topics}.
\bjtitle{Nature Communications}
\bvolume{10}(\bissue{1}),
\bfpage{3439}
(\byear{2019}).
\doiurl{10.1038/s41467-019-11401-8}.
\bcomment{Number: 1 Publisher: Nature Publishing Group}.
Accessed 2023-02-21
\end{barticle}
\endbibitem

\bibitem{aleta_explore_2019}
\begin{barticle}
\bauthor{\bsnm{Aleta}, \binits{A.}},
\bauthor{\bsnm{Meloni}, \binits{S.}},
\bauthor{\bsnm{Perra}, \binits{N.}},
\bauthor{\bsnm{Moreno}, \binits{Y.}}:
\batitle{Explore with caution: mapping the evolution of scientific interest in
  physics}.
\bjtitle{EPJ Data Science}
\bvolume{8}(\bissue{1}),
\bfpage{1}--\blpage{15}
(\byear{2019}).
\doiurl{10.1140/epjds/s13688-019-0205-9}.
\bcomment{Number: 1 Publisher: SpringerOpen}.
Accessed 2023-02-21
\end{barticle}
\endbibitem

\bibitem{refId0}
\begin{barticle}
\bauthor{\bsnm{{Tuninetti, Marta}}},
\bauthor{\bsnm{{Aleta, Alberto}}},
\bauthor{\bsnm{{Paolotti, Daniela}}},
\bauthor{\bsnm{{Moreno, Yamir}}},
\bauthor{\bsnm{{Starnini, Michele}}}:
\batitle{Prediction of new scientific collaborations through multiplex
  networks}.
\bjtitle{EPJ Data Sci.}
\bvolume{10}(\bissue{1}),
\bfpage{25}
(\byear{2021}).
\doiurl{10.1140/epjds/s13688-021-00282-x}
\end{barticle}
\endbibitem

\bibitem{holden_federation_1974}
\begin{barticle}
\bauthor{\bsnm{Holden}, \binits{C.}}:
\batitle{Federation of {Scientists} {Plans} "{Great} {Leap} {Forward}"}.
\bjtitle{Science}
\bvolume{185}(\bissue{4145}),
\bfpage{47}--\blpage{47}
(\byear{1974}).
\doiurl{10.1126/science.185.4145.47.a}.
\bcomment{Publisher: American Association for the Advancement of Science}.
Accessed 2022-12-21
\end{barticle}
\endbibitem

\bibitem{wu_understanding_2021}
\begin{barticle}
\bauthor{\bsnm{Wu}, \binits{L.}},
\bauthor{\bsnm{Hasan}, \binits{S.}},
\bauthor{\bsnm{Chung}, \binits{Y.}},
\bauthor{\bsnm{Kang}, \binits{J.E.}}:
\batitle{Understanding the {Heterogeneity} of {Human} {Mobility} {Patterns}:
  {User} {Characteristics} and {Modal} {Preferences}}.
\bjtitle{Sustainability}
\bvolume{13}(\bissue{24}),
\bfpage{13921}
(\byear{2021}).
\doiurl{10.3390/su132413921}.
\bcomment{Number: 24 Publisher: Multidisciplinary Digital Publishing Institute}
\end{barticle}
\endbibitem

\bibitem{ubaldi_heterogeneity_2021}
\begin{botherref}
\oauthor{\bsnm{Ubaldi}, \binits{E.}},
\oauthor{\bsnm{Monechi}, \binits{B.}},
\oauthor{\bsnm{Chiappetta}, \binits{C.}},
\oauthor{\bsnm{Loreto}, \binits{V.}}:
Heterogeneity and segregation of mobility patterns.
Handbook on Entropy, Complexity and Spatial Dynamics,
486--509
(2021).
ISBN: 9781839100598 Publisher: Edward Elgar Publishing Section: Handbook on
  Entropy, Complexity and Spatial Dynamics
\end{botherref}
\endbibitem

\bibitem{barbosa_human_2018-1}
\begin{barticle}
\bauthor{\bsnm{Barbosa}, \binits{H.}},
\bauthor{\bsnm{Barthelemy}, \binits{M.}},
\bauthor{\bsnm{Ghoshal}, \binits{G.}},
\bauthor{\bsnm{James}, \binits{C.R.}},
\bauthor{\bsnm{Lenormand}, \binits{M.}},
\bauthor{\bsnm{Louail}, \binits{T.}},
\bauthor{\bsnm{Menezes}, \binits{R.}},
\bauthor{\bsnm{Ramasco}, \binits{J.J.}},
\bauthor{\bsnm{Simini}, \binits{F.}},
\bauthor{\bsnm{Tomasini}, \binits{M.}}:
\batitle{Human mobility: {Models} and applications}.
\bjtitle{Physics Reports}
\bvolume{734},
\bfpage{1}--\blpage{74}
(\byear{2018}).
\doiurl{10.1016/j.physrep.2018.01.001}
\end{barticle}
\endbibitem

\bibitem{schlapfer2021universal}
\begin{barticle}
\bauthor{\bsnm{Schl{\"a}pfer}, \binits{M.}},
\bauthor{\bsnm{Dong}, \binits{L.}},
\bauthor{\bsnm{O’Keeffe}, \binits{K.}},
\bauthor{\bsnm{Santi}, \binits{P.}},
\bauthor{\bsnm{Szell}, \binits{M.}},
\bauthor{\bsnm{Salat}, \binits{H.}},
\bauthor{\bsnm{Anklesaria}, \binits{S.}},
\bauthor{\bsnm{Vazifeh}, \binits{M.}},
\bauthor{\bsnm{Ratti}, \binits{C.}},
\bauthor{\bsnm{West}, \binits{G.B.}}:
\batitle{The universal visitation law of human mobility}.
\bjtitle{Nature}
\bvolume{593}(\bissue{7860}),
\bfpage{522}--\blpage{527}
(\byear{2021})
\end{barticle}
\endbibitem

\bibitem{pappalardo2015returners}
\begin{barticle}
\bauthor{\bsnm{Pappalardo}, \binits{L.}},
\bauthor{\bsnm{Simini}, \binits{F.}},
\bauthor{\bsnm{Rinzivillo}, \binits{S.}},
\bauthor{\bsnm{Pedreschi}, \binits{D.}},
\bauthor{\bsnm{Giannotti}, \binits{F.}},
\bauthor{\bsnm{Barab{\'a}si}, \binits{A.-L.}}:
\batitle{Returners and explorers dichotomy in human mobility}.
\bjtitle{Nature communications}
\bvolume{6}(\bissue{1}),
\bfpage{1}--\blpage{8}
(\byear{2015})
\end{barticle}
\endbibitem

\bibitem{simini2021deep}
\begin{barticle}
\bauthor{\bsnm{Simini}, \binits{F.}},
\bauthor{\bsnm{Barlacchi}, \binits{G.}},
\bauthor{\bsnm{Luca}, \binits{M.}},
\bauthor{\bsnm{Pappalardo}, \binits{L.}}:
\batitle{A deep gravity model for mobility flows generation}.
\bjtitle{Nature communications}
\bvolume{12}(\bissue{1}),
\bfpage{1}--\blpage{13}
(\byear{2021})
\end{barticle}
\endbibitem

\bibitem{alessandretti2020scales}
\begin{barticle}
\bauthor{\bsnm{Alessandretti}, \binits{L.}},
\bauthor{\bsnm{Aslak}, \binits{U.}},
\bauthor{\bsnm{Lehmann}, \binits{S.}}:
\batitle{The scales of human mobility}.
\bjtitle{Nature}
\bvolume{587}(\bissue{7834}),
\bfpage{402}--\blpage{407}
(\byear{2020})
\end{barticle}
\endbibitem

\bibitem{schneider2013unravelling}
\begin{barticle}
\bauthor{\bsnm{Schneider}, \binits{C.M.}},
\bauthor{\bsnm{Belik}, \binits{V.}},
\bauthor{\bsnm{Couronn{\'e}}, \binits{T.}},
\bauthor{\bsnm{Smoreda}, \binits{Z.}},
\bauthor{\bsnm{Gonz{\'a}lez}, \binits{M.C.}}:
\batitle{Unravelling daily human mobility motifs}.
\bjtitle{Journal of The Royal Society Interface}
\bvolume{10}(\bissue{84}),
\bfpage{20130246}
(\byear{2013})
\end{barticle}
\endbibitem

\bibitem{simini2012universal}
\begin{barticle}
\bauthor{\bsnm{Simini}, \binits{F.}},
\bauthor{\bsnm{Gonz{\'a}lez}, \binits{M.C.}},
\bauthor{\bsnm{Maritan}, \binits{A.}},
\bauthor{\bsnm{Barab{\'a}si}, \binits{A.-L.}}:
\batitle{A universal model for mobility and migration patterns}.
\bjtitle{Nature}
\bvolume{484}(\bissue{7392}),
\bfpage{96}--\blpage{100}
(\byear{2012})
\end{barticle}
\endbibitem

\bibitem{wilson_statistical_1967}
\begin{barticle}
\bauthor{\bsnm{Wilson}, \binits{A.G.}}:
\batitle{A statistical theory of spatial distribution models}.
\bjtitle{Transportation Research}
\bvolume{1}(\bissue{3}),
\bfpage{253}--\blpage{269}
(\byear{1967}).
\doiurl{10.1016/0041-1647(67)90035-4}
\end{barticle}
\endbibitem

\bibitem{hills2015exploration}
\begin{barticle}
\bauthor{\bsnm{Hills}, \binits{T.T.}},
\bauthor{\bsnm{Todd}, \binits{P.M.}},
\bauthor{\bsnm{Lazer}, \binits{D.}},
\bauthor{\bsnm{Redish}, \binits{A.D.}},
\bauthor{\bsnm{Couzin}, \binits{I.D.}},
\bauthor{\bsnm{Group}, \binits{C.S.R.}}, \betal:
\batitle{Exploration versus exploitation in space, mind, and society}.
\bjtitle{Trends in cognitive sciences}
\bvolume{19}(\bissue{1}),
\bfpage{46}--\blpage{54}
(\byear{2015})
\end{barticle}
\endbibitem

\bibitem{hills2006animal}
\begin{barticle}
\bauthor{\bsnm{Hills}, \binits{T.T.}}:
\batitle{Animal foraging and the evolution of goal-directed cognition}.
\bjtitle{Cognitive science}
\bvolume{30}(\bissue{1}),
\bfpage{3}--\blpage{41}
(\byear{2006})
\end{barticle}
\endbibitem

\bibitem{bonabeau1999swarm}
\begin{botherref}
\oauthor{\bsnm{Bonabeau}, \binits{E.}},
\oauthor{\bsnm{Dorigo}, \binits{M.}},
\oauthor{\bsnm{Theraulaz}, \binits{G.}},
\oauthor{\bsnm{Theraulaz}, \binits{G.}}:
Swarm intelligence: from natural to artificial systems
(1)
(1999)
\end{botherref}
\endbibitem

\bibitem{hills2012optimal}
\begin{barticle}
\bauthor{\bsnm{Hills}, \binits{T.T.}},
\bauthor{\bsnm{Jones}, \binits{M.N.}},
\bauthor{\bsnm{Todd}, \binits{P.M.}}:
\batitle{Optimal foraging in semantic memory.}
\bjtitle{Psychological review}
\bvolume{119}(\bissue{2}),
\bfpage{431}
(\byear{2012})
\end{barticle}
\endbibitem

\bibitem{march_exploration_1991}
\begin{barticle}
\bauthor{\bsnm{March}, \binits{J.G.}}:
\batitle{Exploration and {Exploitation} in {Organizational} {Learning}}.
\bjtitle{Organization Science}
\bvolume{2}(\bissue{1}),
\bfpage{71}--\blpage{87}
(\byear{1991}).
\bcomment{Publisher: INFORMS}
\end{barticle}
\endbibitem

\bibitem{zhao2014scaling}
\begin{barticle}
\bauthor{\bsnm{Zhao}, \binits{Z.-D.}},
\bauthor{\bsnm{Huang}, \binits{Z.-G.}},
\bauthor{\bsnm{Huang}, \binits{L.}},
\bauthor{\bsnm{Liu}, \binits{H.}},
\bauthor{\bsnm{Lai}, \binits{Y.-C.}}:
\batitle{Scaling and correlation of human movements in cyberspace and physical
  space}.
\bjtitle{Physical Review E}
\bvolume{90}(\bissue{5}),
\bfpage{050802}
(\byear{2014})
\end{barticle}
\endbibitem

\bibitem{hu2018life}
\begin{bchapter}
\bauthor{\bsnm{Hu}, \binits{T.}},
\bauthor{\bsnm{Luo}, \binits{J.}},
\bauthor{\bsnm{Liu}, \binits{W.}}:
\bctitle{Life in the" matrix": Human mobility patterns in the cyber space}.
In: \bbtitle{Twelfth International AAAI Conference on Web and Social Media}
(\byear{2018})
\end{bchapter}
\endbibitem

\bibitem{barbosa2016returners}
\begin{bbook}
\bauthor{\bsnm{Barbosa}, \binits{H.S.}},
\bauthor{\bparticle{de} \bsnm{Lima~Neto}, \binits{F.B.}},
\bauthor{\bsnm{Evsukoff}, \binits{A.}},
\bauthor{\bsnm{Menezes}, \binits{R.}}:
\bbtitle{Returners and Explorers Dichotomy in Web Browsing Behavior---A Human
  Mobility Approach},
pp. \bfpage{173}--\blpage{184}.
\bpublisher{Springer},
\blocation{Cham}
(\byear{2016}).
\doiurl{10.1007/978-3-319-30569-1_13}
\end{bbook}
\endbibitem

\bibitem{milojevic2015quantifying}
\begin{barticle}
\bauthor{\bsnm{Milojevi{\'c}}, \binits{S.}}:
\batitle{Quantifying the cognitive extent of science}.
\bjtitle{Journal of Informetrics}
\bvolume{9}(\bissue{4}),
\bfpage{962}--\blpage{973}
(\byear{2015})
\end{barticle}
\endbibitem

\bibitem{milojevic_cognitive_2011}
\begin{barticle}
\bauthor{\bsnm{Milojević}, \binits{S.}},
\bauthor{\bsnm{Sugimoto}, \binits{C.R.}},
\bauthor{\bsnm{Yan}, \binits{E.}},
\bauthor{\bsnm{Ding}, \binits{Y.}}:
\batitle{The cognitive structure of {Library} and {Information} {Science}:
  {Analysis} of article title words}.
\bjtitle{Journal of the American Society for Information Science and
  Technology}
\bvolume{62}(\bissue{10}),
\bfpage{1933}--\blpage{1953}
(\byear{2011}).
\doiurl{10.1002/asi.21602}
\end{barticle}
\endbibitem

\bibitem{peng2021neural}
\begin{barticle}
\bauthor{\bsnm{Peng}, \binits{H.}},
\bauthor{\bsnm{Ke}, \binits{Q.}},
\bauthor{\bsnm{Budak}, \binits{C.}},
\bauthor{\bsnm{Romero}, \binits{D.M.}},
\bauthor{\bsnm{Ahn}, \binits{Y.-Y.}}:
\batitle{Neural embeddings of scholarly periodicals reveal complex disciplinary
  organizations}.
\bjtitle{Science Advances}
\bvolume{7}(\bissue{17}),
\bfpage{9004}
(\byear{2021})
\end{barticle}
\endbibitem

\bibitem{ying_modeling_2015}
\begin{botherref}
\oauthor{\bsnm{Ying}, \binits{Q.F.}},
\oauthor{\bsnm{Venkatramanan}, \binits{S.}},
\oauthor{\bsnm{Chiu}, \binits{D.M.}}:
Modeling and {Analysis} of {Scholar} {Mobility} on {Scientific} {Landscape}.
arXiv.
arXiv:1502.00523 [physics]
(2015).
\url{http://arxiv.org/abs/1502.00523}
\end{botherref}
\endbibitem

\bibitem{singh2022quantifying}
\begin{barticle}
\bauthor{\bsnm{Singh}, \binits{C.K.}},
\bauthor{\bsnm{Barme}, \binits{E.}},
\bauthor{\bsnm{Ward}, \binits{R.}},
\bauthor{\bsnm{Tupikina}, \binits{L.}},
\bauthor{\bsnm{Santolini}, \binits{M.}}:
\batitle{Quantifying the rise and fall of scientific fields}.
\bjtitle{PloS one}
\bvolume{17}(\bissue{6}),
\bfpage{0270131}
(\byear{2022})
\end{barticle}
\endbibitem

\bibitem{van2008visualizing}
\begin{botherref}
\oauthor{\bparticle{Van~der} \bsnm{Maaten}, \binits{L.}},
\oauthor{\bsnm{Hinton}, \binits{G.}}:
Visualizing data using t-sne.
Journal of machine learning research
\textbf{9}(11)
(2008)
\end{botherref}
\endbibitem

\bibitem{van2009learning}
\begin{bchapter}
\bauthor{\bsnm{Van Der~Maaten}, \binits{L.}}:
\bctitle{Learning a parametric embedding by preserving local structure}.
In: \bbtitle{Artificial Intelligence and Statistics},
pp. \bfpage{384}--\blpage{391}
(\byear{2009}).
\bcomment{PMLR}
\end{bchapter}
\endbibitem

\bibitem{WILSON1967253}
\begin{barticle}
\bauthor{\bsnm{Wilson}, \binits{A.G.}}:
\batitle{A statistical theory of spatial distribution models}.
\bjtitle{Transportation Research}
\bvolume{1}(\bissue{3}),
\bfpage{253}--\blpage{269}
(\byear{1967}).
\doiurl{10.1016/0041-1647(67)90035-4}
\end{barticle}
\endbibitem

\bibitem{senior1979gravity}
\begin{barticle}
\bauthor{\bsnm{Senior}, \binits{M.L.}}:
\batitle{From gravity modelling to entropy maximizing: a pedagogic guide}.
\bjtitle{Progress in Human Geography}
\bvolume{3}(\bissue{2}),
\bfpage{175}--\blpage{210}
(\byear{1979})
\end{barticle}
\endbibitem

\bibitem{wilson2013entropy}
\begin{bbook}
\bauthor{\bsnm{Wilson}, \binits{A.}}:
\bbtitle{Entropy in Urban and Regional Modelling (Routledge Revivals).
  Routledge},
(\byear{2013})
\end{bbook}
\endbibitem

\bibitem{ribeiro_mathematical_2021}
\begin{botherref}
\oauthor{\bsnm{Ribeiro}, \binits{F.L.}},
\oauthor{\bsnm{Rybski}, \binits{D.}}:
Mathematical models to explain the origin of urban scaling laws: a synthetic
  review.
arXiv.
arXiv:2111.08365 [physics]
(2021).
\doiurl{10.48550/arXiv.2111.08365}.
\url{http://arxiv.org/abs/2111.08365}
Accessed 2022-12-26
\end{botherref}
\endbibitem

\bibitem{klafter2011first}
\begin{botherref}
\oauthor{\bsnm{Klafter}, \binits{J.}},
\oauthor{\bsnm{Sokolov}, \binits{I.M.}}:
First steps in random walks: From tools to applications
(2011).
\doiurl{10.1093/acprof:oso/9780199234868.001.0001}
\end{botherref}
\endbibitem

\bibitem{cui_aging_2022}
\begin{botherref}
\oauthor{\bsnm{Cui}, \binits{H.}},
\oauthor{\bsnm{Wu}, \binits{L.}},
\oauthor{\bsnm{Evans}, \binits{J.A.}}:
Aging {Scientists} and {Slowed} {Advance}.
arXiv.
arXiv:2202.04044 [cs]
(2022).
\doiurl{10.48550/arXiv.2202.04044}.
\url{http://arxiv.org/abs/2202.04044}
Accessed 2023-02-21
\end{botherref}
\endbibitem

\bibitem{park_papers_2023}
\begin{barticle}
\bauthor{\bsnm{Park}, \binits{M.}},
\bauthor{\bsnm{Leahey}, \binits{E.}},
\bauthor{\bsnm{Funk}, \binits{R.J.}}:
\batitle{Papers and patents are becoming less disruptive over time}.
\bjtitle{Nature}
\bvolume{613}(\bissue{7942}),
\bfpage{138}--\blpage{144}
(\byear{2023}).
\doiurl{10.1038/s41586-022-05543-x}.
Accessed 2023-01-13
\end{barticle}
\endbibitem

\bibitem{clement_use_2019}
\begin{botherref}
\oauthor{\bsnm{Clement}, \binits{C.B.}},
\oauthor{\bsnm{Bierbaum}, \binits{M.}},
\oauthor{\bsnm{O'Keeffe}, \binits{K.P.}},
\oauthor{\bsnm{Alemi}, \binits{A.A.}}:
On the {Use} of {ArXiv} as a {Dataset}.
arXiv.
arXiv:1905.00075 [physics]
(2019).
\doiurl{10.48550/arXiv.1905.00075}.
\url{http://arxiv.org/abs/1905.00075}
Accessed 2022-12-23
\end{botherref}
\endbibitem

\bibitem{kuhn_essential_1979}
\begin{bbook}
\bauthor{\bsnm{Kuhn}, \binits{T.S.}}:
\bbtitle{The {Essential} {Tension}: {Selected} {Studies} in {Scientific}
  {Tradition} and {Change}}.
\bpublisher{University of Chicago Press},
\blocation{Chicago, IL}
(\byear{1979}).
\burl{https://press.uchicago.edu/ucp/books/book/chicago/E/bo5970650.html}
Accessed 2023-02-21
\end{bbook}
\endbibitem

\bibitem{lanoiselee_unraveling_2017}
\begin{barticle}
\bauthor{\bsnm{Lanoiselée}, \binits{Y.}},
\bauthor{\bsnm{Grebenkov}, \binits{D.S.}}:
\batitle{Unraveling intermittent features in single-particle trajectories by a
  local convex hull method}.
\bjtitle{Physical Review E}
\bvolume{96}(\bissue{2}),
\bfpage{022144}
(\byear{2017}).
\doiurl{10.1103/PhysRevE.96.022144}.
\bcomment{Publisher: American Physical Society}.
Accessed 2022-12-21
\end{barticle}
\endbibitem

\bibitem{beltagy_scibert_2019}
\begin{bchapter}
\bauthor{\bsnm{Beltagy}, \binits{I.}},
\bauthor{\bsnm{Lo}, \binits{K.}},
\bauthor{\bsnm{Cohan}, \binits{A.}}:
\bctitle{{SciBERT}: {A} {Pretrained} {Language} {Model} for {Scientific}
  {Text}}.
In: \bbtitle{Proceedings of the 2019 {Conference} on {Empirical} {Methods} in
  {Natural} {Language} {Processing} and the 9th {International} {Joint}
  {Conference} on {Natural} {Language} {Processing} ({EMNLP}-{IJCNLP})},
pp. \bfpage{3615}--\blpage{3620}.
\bpublisher{Association for Computational Linguistics},
\blocation{Hong Kong, China}
(\byear{2019}).
\doiurl{10.18653/v1/D19-1371}.
\burl{https://aclanthology.org/D19-1371}
Accessed 2023-02-20
\end{bchapter}
\endbibitem

\bibitem{wuchty2007increasing}
\begin{barticle}
\bauthor{\bsnm{Wuchty}, \binits{S.}},
\bauthor{\bsnm{Jones}, \binits{B.F.}},
\bauthor{\bsnm{Uzzi}, \binits{B.}}:
\batitle{The increasing dominance of teams in production of knowledge}.
\bjtitle{Science}
\bvolume{316}(\bissue{5827}),
\bfpage{1036}--\blpage{1039}
(\byear{2007})
\end{barticle}
\endbibitem

\bibitem{sekara2018chaperone}
\begin{barticle}
\bauthor{\bsnm{Sekara}, \binits{V.}},
\bauthor{\bsnm{Deville}, \binits{P.}},
\bauthor{\bsnm{Ahnert}, \binits{S.E.}},
\bauthor{\bsnm{Barab{\'a}si}, \binits{A.-L.}},
\bauthor{\bsnm{Sinatra}, \binits{R.}},
\bauthor{\bsnm{Lehmann}, \binits{S.}}:
\batitle{The chaperone effect in scientific publishing}.
\bjtitle{Proceedings of the National Academy of Sciences}
\bvolume{115}(\bissue{50}),
\bfpage{12603}--\blpage{12607}
(\byear{2018})
\end{barticle}
\endbibitem

\bibitem{pedregosa_scikit-learn_2011}
\begin{barticle}
\bauthor{\bsnm{Pedregosa}, \binits{F.}},
\bauthor{\bsnm{Varoquaux}, \binits{G.}},
\bauthor{\bsnm{Gramfort}, \binits{A.}},
\bauthor{\bsnm{Michel}, \binits{V.}},
\bauthor{\bsnm{Thirion}, \binits{B.}},
\bauthor{\bsnm{Grisel}, \binits{O.}},
\bauthor{\bsnm{Blondel}, \binits{M.}},
\bauthor{\bsnm{Prettenhofer}, \binits{P.}},
\bauthor{\bsnm{Weiss}, \binits{R.}},
\bauthor{\bsnm{Dubourg}, \binits{V.}},
\bauthor{\bsnm{Vanderplas}, \binits{J.}},
\bauthor{\bsnm{Passos}, \binits{A.}},
\bauthor{\bsnm{Cournapeau}, \binits{D.}},
\bauthor{\bsnm{Brucher}, \binits{M.}},
\bauthor{\bsnm{Perrot}, \binits{M.}},
\bauthor{\bsnm{Duchesnay}, \binits{E.}}:
\batitle{Scikit-learn: {Machine} {Learning} in {Python}}.
\bjtitle{The Journal of Machine Learning Research}
\bvolume{12}(\bissue{null}),
\bfpage{2825}--\blpage{2830}
(\byear{2011})
\end{barticle}
\endbibitem

\bibitem{wang2021understanding}
\begin{barticle}
\bauthor{\bsnm{Wang}, \binits{Y.}},
\bauthor{\bsnm{Huang}, \binits{H.}},
\bauthor{\bsnm{Rudin}, \binits{C.}},
\bauthor{\bsnm{Shaposhnik}, \binits{Y.}}:
\batitle{Understanding how dimension reduction tools work: An empirical
  approach to deciphering t-sne, umap, trimap, and pacmap for data
  visualization.}
\bjtitle{J. Mach. Learn. Res.}
\bvolume{22}(\bissue{201}),
\bfpage{1}--\blpage{73}
(\byear{2021})
\end{barticle}
\endbibitem

\bibitem{mcinnes2018umap}
\begin{botherref}
\oauthor{\bsnm{McInnes}, \binits{L.}},
\oauthor{\bsnm{Healy}, \binits{J.}},
\oauthor{\bsnm{Melville}, \binits{J.}}:
Umap: Uniform manifold approximation and projection for dimension reduction.
arXiv preprint arXiv:1802.03426
(2018)
\end{botherref}
\endbibitem

\bibitem{wu_large_2019}
\begin{barticle}
\bauthor{\bsnm{Wu}, \binits{L.}},
\bauthor{\bsnm{Wang}, \binits{D.}},
\bauthor{\bsnm{Evans}, \binits{J.A.}}:
\batitle{Large teams develop and small teams disrupt science and technology}.
\bjtitle{Nature}
\bvolume{566}(\bissue{7744}),
\bfpage{378}--\blpage{382}
(\byear{2019}).
\doiurl{10.1038/s41586-019-0941-9}.
\bcomment{Number: 7744 Publisher: Nature Publishing Group}.
Accessed 2021-03-19
\end{barticle}
\endbibitem

\end{thebibliography}


\backmatter

\end{document}